\documentclass[fleqn,usenatbib]{mnras}

\usepackage[T1]{fontenc}
\usepackage{ae,aecompl}

\usepackage{graphicx}	% Including figure files
\usepackage{amsmath}	% Advanced maths commands
\usepackage{amssymb}	% Extra maths symbols

\title[Spiral structure of M33]
{
Simulations of the flocculent spiral M33: what drives the spiral structure?
}
\author[Dobbs]
{C. L. Dobbs\thanks{E-mail:
dobbs@astro.ex.ac.uk}$^{1}$, A. R. Pettitt$^{2}$, E. Corbelli$^{3}$ and J. E. Pringle$^4$\\
$^1$ School of Physics and Astronomy, University of Exeter, Stocker Road, Exeter, EX4 4QL, UK \\
$^{2}$ Department of Physics, Faculty of Science, Hokkaido University, Sapporo 060-0810, Japan\\
$^{3}$ INAF-Osservatorio Astrofisico di Arcetri, Largo E. Fermi 5, 50125 Firenze, Italy \\
$^{4}$ Institute of Astronomy, Madingley Road, Cambridge, CB3 0HA}
\def\aap{{ A\&A}}

\def\apjl{{ApJL}}

\def\apjs{{ApJS}}

\begin{document}
\label{firstpage}
\date{\today}

\pagerange{\pageref{firstpage}--\pageref{lastpage}} \pubyear{2012}

\maketitle

\begin{abstract}
We perform simulations of isolated galaxies in order to investigate the likely origin of the spiral structure in M33. In our models, we find that gravitational instabilities in the stars and gas are able to reproduce the observed spiral pattern and velocity field of M33, as seen in HI, and no interaction is required. We also find that the optimum models have high levels of stellar feedback which create large holes similar to those observed in M33, whilst lower levels of feedback tend to produce a large amount of small scale structure, and undisturbed long filaments of high surface density gas, hardly detected in the M33 disc. The gas component appears to have a significant role in producing the structure, so if there is little feedback, both the gas and stars organise into clear spiral  arms, likely due to a lower combined $Q$ (using gas and stars), and the ready ability of cold gas to undergo spiral shocks. By contrast models with higher feedback have weaker spiral structure, especially in the stellar component, compared to grand design galaxies. We did not see a large difference in the behaviour of $Q_{stars}$ with most of these models, however, because $Q_{stars}$ stayed relatively constant unless the disc was more strongly unstable. Our models suggest that although the stars produce some underlying spiral structure, this is relatively weak, and the gas physics has a considerable role in producing the large scale structure of the ISM in flocculent spirals.
\end{abstract}


\section{Introduction}
Star formation is likely dependent on both the large scale dynamics of a galaxy and the smaller scale gas physics. For example, in many galaxies, star formation is associated with the spiral arms. However the relative contributions of spiral structure and gas physics on the overall distribution of the interstellar medium, and locations of star formation, are not yet clear.  In this paper we examine the influence of both of these by studying the origin of spiral arms, and the effect of feedback and gas physics on the large scale spiral structure for the nearby galaxy M33. In particular we address what produces the spiral structure in M33, and the relevance of stellar feedback in reproducing the observed gas distribution. 

There are several mechanisms for spiral arm formation, which include steady state density wave theory, tidal interactions, bars and localised gravitational instabilities (for a review see \citealt{DobbsBaba2014}). Whilst there are many numerical simulations which demonstrate these mechanisms, and the resulting properties of the spiral arms, identifying the mechanism responsible for spiral arms in particular galaxies is still relatively rare. In some cases, galaxy interactions are clearly generating spiral structure. For example, models of the interaction of M51 and its neighbour NGC 5195 have been shown to reproduce the spiral arm pattern very well \citep{Hernquist1990,Salo2000b,Dobbs2010}. Numerical simulations have also produced the structure of the Antennae galaxies as they undergo a merger \citep{Karl2010,Renaud2015b,Lahen2017}. For other galaxies, it is unclear what is producing the spiral arms. In isolated galaxies, gravitational instabilities in the stars may dominate the spiral arm structure. For the Milky Way, galaxy models where spiral arms are generated by gravitational instabilities produce a better match to both GAIA data \citep{Baba2018} and the Galactic CO map compared to models using a fixed spiral arm potential \citep{Baba2010,Pettitt2015}. \citet{Purcell2011} model the interaction of the Milky Way and Sagittarius, suggesting that this generates the spiral arms, though their simulations do not produce a very detailed spiral morphology of the Galaxy.  

A second question is what is the importance of spiral structure for star formation. In some simulations, spiral arms determine the location of the formation of stars and molecular clouds (e.g. \citealt{Dobbs2013, Pettitt2017}), even though the spiral arms might not have a strong effect on the rate of star formation. This is often clearest in models where spiral arms are triggered by interactions \citep{Pettitt2017}, set up as fixed spiral arms \citep{Dobbs2013}, but may also occur for transient spiral arms induced by gravitational instabilities in the stellar disc. Other models do not presume that spiral arms have a large role in star formation. They instead suppose a `supernovae driven ISM', whereby feedback from supernovae produces shells and triggers further star formation \citep{Padoan2011, Gatto2017, deAvillez2005, Inutsuka2015}. The large majority of these models do not include spiral arms. 

Observed galaxies exhibit different strength arms, suggesting that the relative importance of spiral arms, versus other processes such as stellar feedback, may vary. M51 is characterised by massive GMCs and spurs hosting clusters along the spiral arms, thus the strong tidally induced spiral arms appear to play a significant role in star formation and shaping the ISM. In galaxies with particularly strong spiral arms, there may not be so much difference in the spiral structure whether or not stellar feedback occurs (see for example models with different levels of stellar feedback in \citealt{Dobbs2011new}). However in other galaxy types, and / or galaxies with weak spiral structure, stellar feedback may have a greater role in determining the structure of the gas and stars.  

M33 is one of the closest spiral galaxies to us and a member of the Local Group. As such there are high resolution observations in HI, CO, and stellar cluster catalogues \citep{Sharma2011,Gratier2012,Miura2012,Corbelli2014,Corbelli2017,Kam2017}. M33 does not have particularly strong spiral arms or grand design structure. Instead it exhibits a number of weaker spiral arms more characteristic of a flocculent spiral galaxy \citep{Humphreys1980}. Although in the near infra-red (NIR) there are two slightly more prominent arms, the spiral structure is very weak compared to grand design galaxies such as M51, and further spiral arms are still evident \citep{Jarrett2003}.

 There have been few studies to try and examine the structure of the M33 disc and determine its origin. A number of studies have investigated interactions of M33 with other members of the Local Group \citep{Patel2017, Bekki2008,Semczuk2018}, although they did not investigate the detailed structure of the disc of M33. \citet{Patel2017} show that M33 appears to be approaching M31 for the first time. This implies that the current spiral arms are not the result of the interaction with M31, and instead M33 can essentially be considered as an isolated galaxy. \citet{Semczuk2018} instead propose an orbit whereby M33 and M31 had a close encounter 2 Gyr ago which produced two tidal arms. \citet{Rahimi2012} model an isolated `M33 type galaxy' in terms of mass and size, but don't make direct comparisons of their models with the spiral structure of M33. They do however find that a relatively large amount of stellar feedback is required to produce a resemblance to M33.

In this paper, we perform numerical simulations of an isolated galaxy with a stellar and gas disc, and dark matter halo chosen to match the M33 galaxy. We investigate whether gaseous spiral arms resembling those of M33 can be produced from gravitational instabilities in the disc, without an interaction. We also examine the role of gas in the disc, in terms of contributing to gravitational instabilities, whether stellar feedback is important to reproduce the spiral structure, and the dependence on the thermal properties of the gas. We perform simulations with two different codes, \textsc{sphNG} and \textsc{gasoline2}.  

\section{Details of \textsc{sphNG} simulations}
We use the Smoothed Particle Hydrodynamics (SPH) code, developed by \citet{Bate1995} for most of these calculations. The code includes star particles \citep{Dobbs2010}, adaptive softening \citep{PM2007} ISM heating and cooling \citep{Dobbs2008}, and H$_2$ and CO formation \citep{Dobbs2008,Pettitt2014}. Stellar feedback is included using the simple prescription of \citet{Dobbs2011new}, whereby an amount of energy given by
\begin{equation}
E=\frac{\epsilon M 10^{51}}{160 \rm{M}_{\odot}} \rm{ergs}
\end{equation}
is inserted for each star formation event. Here $10^{51}$ ergs is the energy released by one supernova, $M$ is the gas mass within a few smoothing lengths, and $\epsilon$ is an efficiency parameter. The parameter $\epsilon$ controls both the amount of star formation per feedback event (which is $\epsilon M$), and the energy added per feedback event according to Equation 1.  We assume that one massive star forms per 160 M$_{\odot}$ of stars formed.
Unlike previous work, here we use the total gas mass $M$ within a few smoothing lengths to calculate the amount of star formation, rather than the mass of molecular hydrogen. For our models of M33, only a small amount of molecular hydrogen is formed, and indeed the actual M33 is mostly HI, so it seemed more reasonable to use the total mass. Energy is inserted as kinetic and thermal energy. The velocities and temperatures of the gas are chosen according to the snowplough solution for a blast wave. Although nominally feedback is inserted following a model for supernovae, in reality we expect other forms of feedback, which inject the ISM with a similar amount of energy, will occur and take place on timescales shorter than supernovae. As such, our feedback prescription may be more representative of other forms of feedback (winds, ionisation, radiation pressure), which act on shorter timescales. We also ran a model with a short, 5 Myr delay (not shown), but this produced similar results.

In the simulations presented here, we model the gas and stars but include an NFW potential 
\begin{equation}
\psi(r)=-\frac{GM_h f}{r} \ln(1+r/r_h)
\end{equation}
where
\begin{equation}
f=\ln (C+1)-\frac{C}{C+1}
\end{equation}
\citep{Navarro1996} for the dark matter halo. 
We adopt a couple of different sets of parameters, but both are based on the parameters used to fit the rotation curve to the stars, gas and an NFW halo by \citet{Corbelli2014}. There is some uncertainty, and degeneracy in the values of these different parameters in \citet{Corbelli2014}, but as we show in Figure~\ref{fig:velcurvesd}, and similar figures in \citet{Corbelli2014}, it is possible to fairly well match the rotation curve of M33. In our first set of models we take $M_h=4.3\times 10^{11}$ M$_{\odot}$ and $r_h=26$ kpc. We take a value of $C=6$ at the low end of the range of \citet{Corbelli2014}, although we also tested using $C=8$ and did not find much difference. 

\subsection{Set up of M33 initial conditions}
\begin{table*}
\begin{tabular}{c|cc|c|c|c|c|c|c|c|c}
 \hline 
 Run &  No. gas & No. star & Gas particle & Total gas & Total stellar & Feedback & $Q_{stars}$  & Length of \\
 & particles & particles & mass (M$_{\odot}$) & mass ($10^9$ M$_{\odot}$) & mass ($10^9$ M$_{\odot}$) & efficiency $\epsilon$ (\%) & & simulation (Myr) \\
 \hline
\textbf{\textsc{sphNG}} & & & & & & & \\
LowF & 1375000 & 1375000 & 2000 & 2.75 & 5.5 & 1 & 1 & 425 \\
MedF & 1375000 & 1375000 & 2000 & 2.75 & 5.5 &  5 & 1 & 675 \\
HighF & 1375000 & 1375000 & 2000 & 2.75 & 5.5 &  20 & 1 &675 \\ 
Static & 1375000 & 1375000 & 2000 &2.75 & 5.5 &  5 & $\infty$ & 425  \\ 
LowQ & 1375000 & 1375000 & 2000 & 2.75 & 5.5 &  5 & 0.7 &  310 \\
StarsOnly & 0 & 1375000 & - & - & 5.5 & - & 1 &  1000 \\
StarsOnlyLowQ & 0 & 1375000 & - & - & 5.5 &  - & 0.7 & 425 \\
MedRes & 2750000 & 2750000 & 1000 & 2.75 & 5.5 & 20 & 1 &425 \\ 
HighRres & 7087500 & 1675000 & 390 & 2.77 & 5.5 & 20 & 1 &425 \\
\textbf{\textsc{gasoline2}} & & & & & & & \\
GSLNfb01 &4000000 & 4100000 & 440 & 1.76 & 4.5 & 1 & 1 &1000\\
GSLNfb05 &4000000 & 4100000 & 440 & 1.76 & 4.5 & 5 & 1 &1000\\
GSLNfb10 &4000000 & 4100000 & 440 & 1.76 & 4.5 & 10 & 1 &1000\\
GSLNfb20 &4000000 & 4100000 & 440 & 1.76 & 4.5 & 20 & 1 &1000\\
\hline
\end{tabular}
\caption{List of calculations performed. By comparison, the observed total gas mass within the optical disc of M33 is around 2 $\times 10^9$~M$_{\odot}$, with an uncertainty of 10--20\%. The stellar mass is around double this, 4.3 $\times 10^9$ M$_{\odot}$, with similar uncertainties \citep{Corbelli2014}.}\label{tab:simtable1}
\end{table*}
We tried a number of different approaches to setting up the initial conditions for our M33 models. 
We initially tried using the mkd95 program, part of the NEMO package, as used in \citet{Dobbs2010} for modelling M51 to set up the M33 galaxy. However in order to acquire a rotation curve similar to that observed, a very large halo was required, and the gas and stars produced many very short spiral arms which did not resemble M33. 

\begin{figure}
\centerline{\includegraphics[scale=0.55]{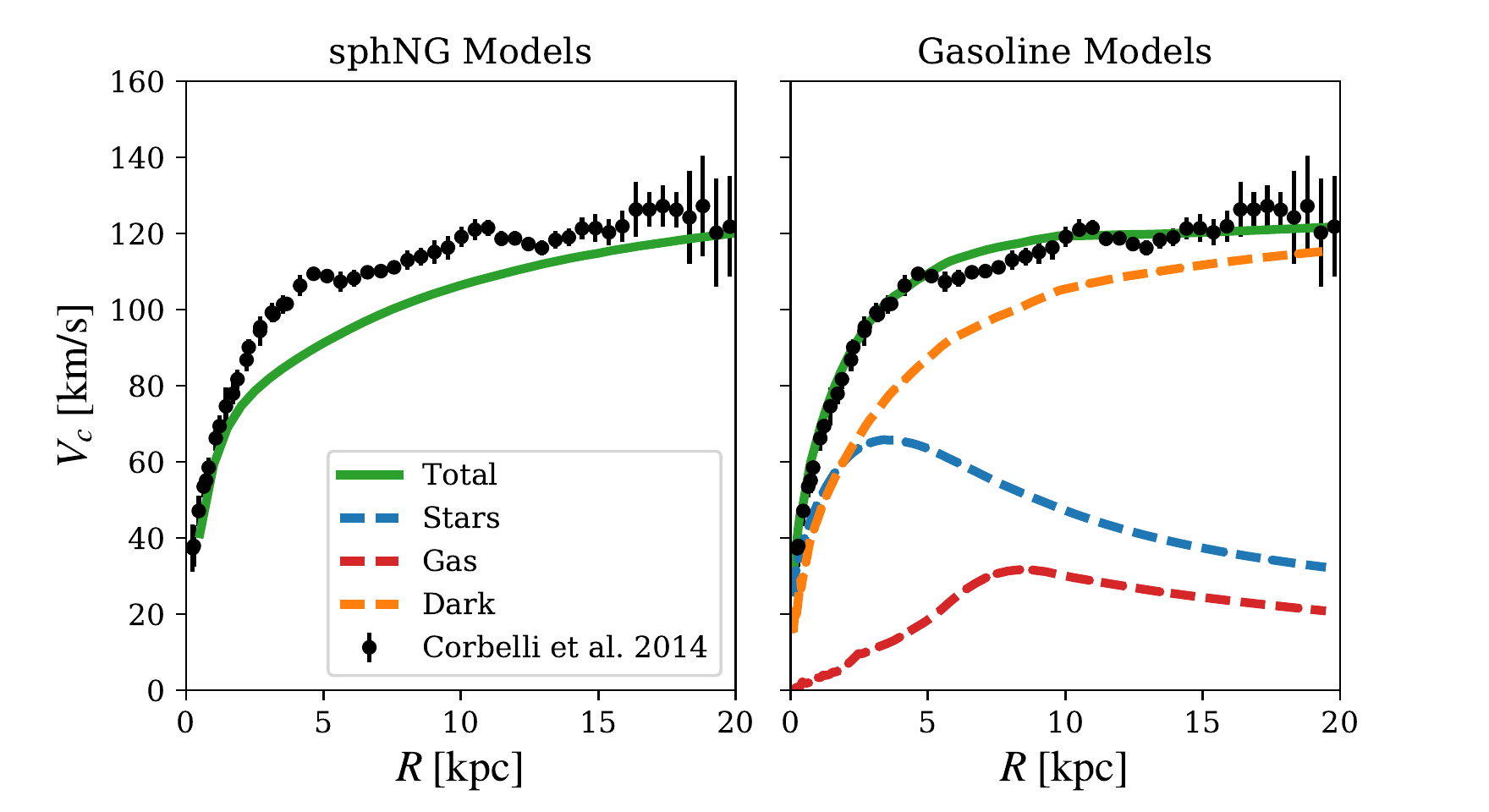}}
\centerline{\includegraphics[scale=0.55]{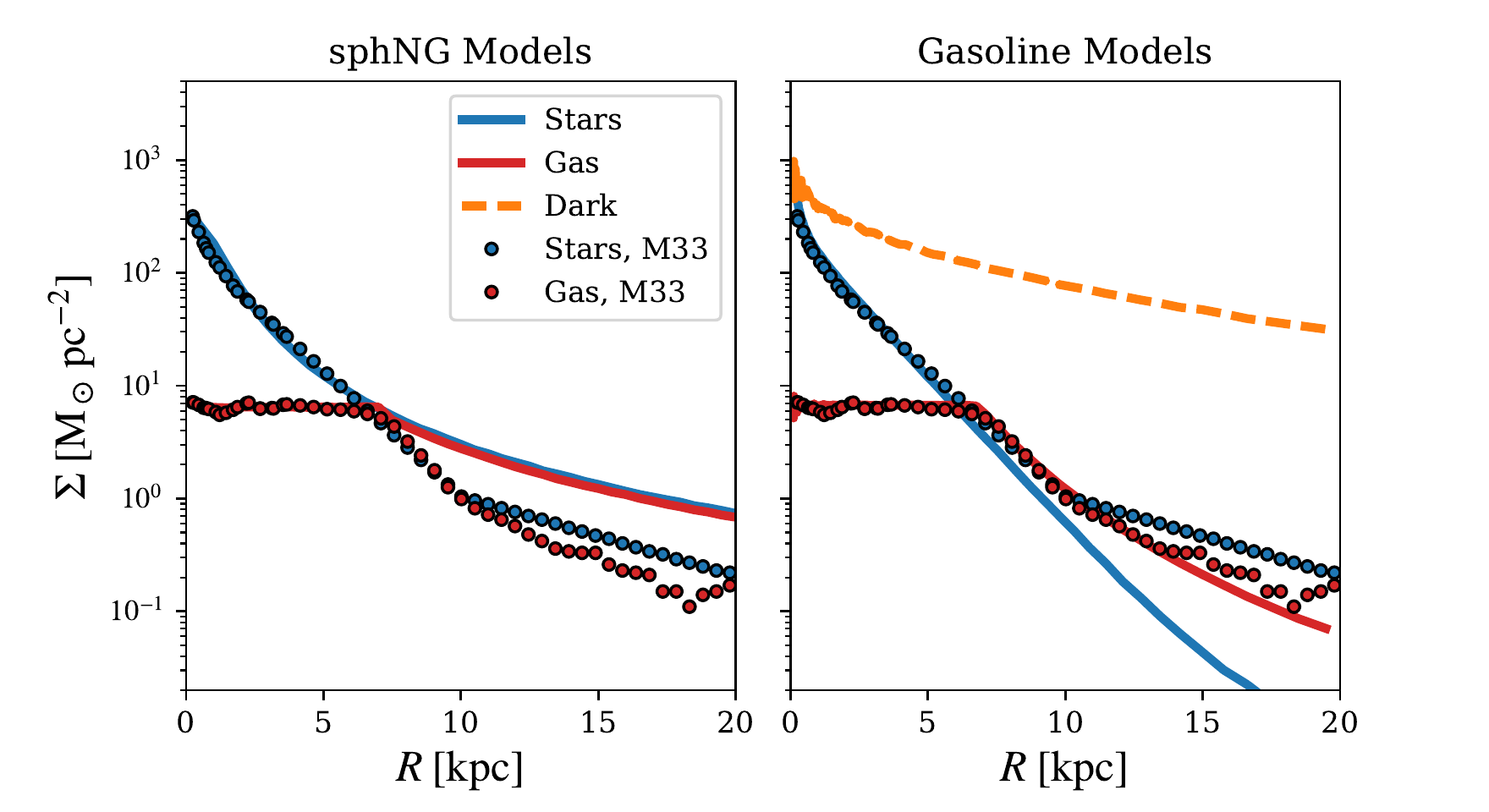}}
\caption{In the top panels we compare the rotation curve for the \textsc{sphNG} and \textsc{gasoline2} models with that of M33. In the lower panels we compare the surface densities of the gas and stars.}
\label{fig:velcurvesd}
\end{figure}

The second approach we used, and which is adopted for the sphNG simulations presented, is to directly allocate gas and star particles according to the observed stellar and gas distributions, from \citet{Corbelli2014}. We set up the stellar mass distribution with a $1/r$ density profile. The gas is distributed uniformly up to a radius of 7 kpc, beyond which the gas also drops off with a $1/r$ profile. We show in Figure~\ref{fig:velcurvesd} the surface densities of the stars and gas for the model, and those measured for the actual M33. Up to a radius of around 7 kpc, the main region of M33 which we are interested in, the gas and stellar surface densities  of the model match M33 well. Beyond this the gas density falls off.  For simplicity, we have continued the same stellar distribution to lower radii (for M33 the stellar profile is also simply an extrapolation at large radii), and equate the gas surface density to the the stellar density, although that means the gas surface density is a little high in the outer parts of the disc. 
This approach does not set up the stars and gas in equilibrium, but has the advantage that there is more flexibility in how to set up the gas and stars.  In particular many schemes to set up stellar discs assume an exponential profile, whereas the M33 stellar disc is not fitted so well by an exponential with a single length scale. We truncate the disc at 20 kpc. We add a constant velocity dispersion of 10 km s$^{-1}$ to the gas, again to agree roughly with observations (there is no evident radial gradient in the dispersion). We set up the velocity dispersions of the stars so that the vertical component of the dispersion matches the vertical gravity \citep{vanderKruit1988}.  This set up a model with $Q\sim1$ for the stars, where
\begin{equation}
Q=\frac{\kappa \sigma}{3.36 G \Sigma}
\end{equation}
\citep{Toomre1964}, where $\kappa$ is the epicyclic frequency, $\sigma$ is the radial velocity dispersion and $\Sigma$ is the stellar surface density. $Q$ increases slightly with radius compared to setting $Q$ as a constant with this method, but models run with constant $Q$ instead produced relatively similar results. The velocity dispersions are then scaled to produce different values of $Q$. For our models with $Q\sim1$, the velocity dispersion at the centre of the disc is around 20--25 km s$^{-1}$, in agreement with observations of M33 \citep{Corbelli2007}. 

Theoretically, we expect a thin stellar disc to be unstable to gravitational instabilities if $Q\lesssim1$. 
For the case of a real, or simulated disc, the disc obviously has a finite scale height, and the disc will contain stars as well as gas. There are several examples of combined $Q$ parameters which take into account both gas and stars \citep{Wang1994,Romeo2011}. However these tend to assume an isothermal gas, whereas the simulations we perform here exhibit a multiphase ISM. So we do not use a combined $Q$ parameter; our values of $Q$ represent the stellar component as given by Equation 4. For our calculations there is more mass in stars than gas, so $Q_{stars}<Q_{gas}$. However the effect of gas on the stability of the disc is not negligible, as we see in the calculations, and we find $Q_{stars}$ does not appear to represent the complete behaviour of the disc.

For the third approach, we take more care to set up the simulations in equilibrium, using \textsc{gasoline2}. We primarily ran these models as a check to see if the structures produced by the sphNG models are an artefact of the initial conditions. We describe these simulations in Section 3.

In all our models, we only consider the atomic gas component of M33, and compare with HI maps of M33. The main reason for this is that molecular gas formation and evolution cannot be well resolved in our simulations. As shown in \citet{Duarte2015}, when stellar feedback is included in galaxy-scale simulations, little gas reaches or stays at the densities required to become molecular. At smaller scales, where it possible to resolve the detailed structure of the clouds, larger amounts of molecular gas more consistent with observations are produced \citep{Duarte2016,Rey-Raposo2017}. For our M33 models, we do not actively switch off H$_2$ formation in the cooling and chemistry routine in the code, but only a minimal amount of molecular gas is formed. The actual M33 does contain some molecular hydrogen in the centre \citep{Corbelli2014}. In our models we see an increase in gas surface density (by a factor of $\sim2$ at the centre of the galaxy (within  a kpc radius), so there is some indication that our models would have more gas at the centre with time, which could correspond to the molecular distribution of the actual M33. However overall M33 is HI dominated, and we would not expect the molecular gas component to have a significant effect on the development of large scale structure in either the real M33 or our models, so we leave any investigations of molecular gas to future work.

We performed a number of different calculations, varying resolution, stellar feedback, minimum temperature of the gas and $Q$ (Table~1). Because of the computational time taken to run simulations at higher resolution, we run our tests at relatively low resolution, and only perform a couple of simulations at high resolution. In the Static model, we do not evolve the stellar disc, so this effectively mimics a disc with very high Q. The structure in the gas is instead primarily driven by the gas physics, i.e. cooling, and stellar feedback.  We also ran a model with $Q\sim2$, which is not listed in Table~1, but the structure of the gas was not that dissimilar from our model with $Q\sim1$ (MedF), so we instead show the Static model where there are clearer differences. A pressure floor is applied below 300 K (see \citealt{Robertson2008}). In all the calculations, all the gas is initially given a temperature of 1000 K. 

\subsection{Expected number of spiral arms}
\begin{figure}
\centerline{\includegraphics[scale=0.32]{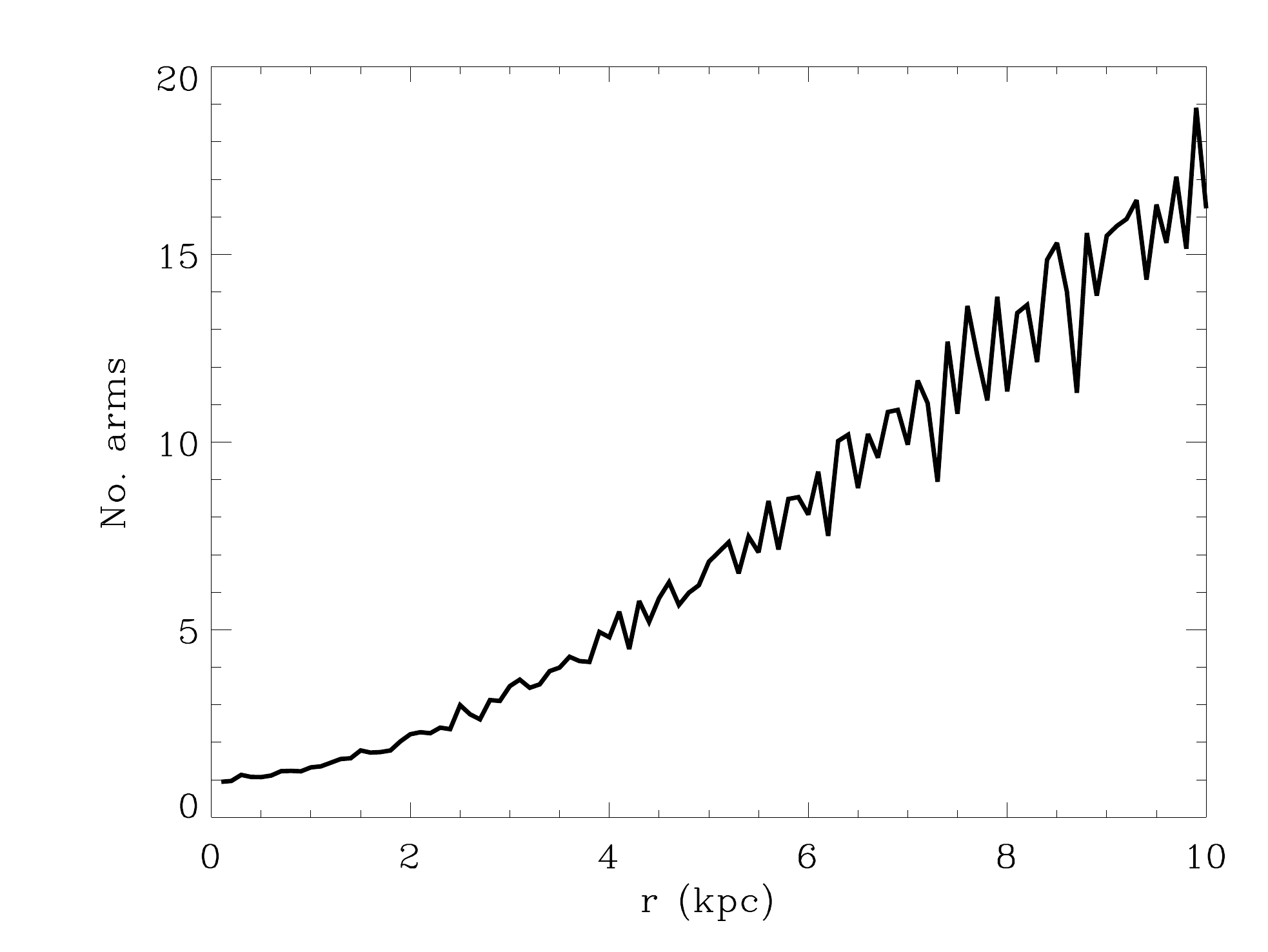}}
\caption{Predicted number of spiral arms for the model MedF.}\label{fig:arms}
\end{figure}

If the number of spiral arms in the galaxy is driven by Toomre instabilities, then the expected number of spiral arms is 
\begin{equation}
m\sim\frac{\kappa^2 R}{4 \pi G \Sigma}
\end{equation}
\citep{Fujii2011,Pettitt2015}. The predicted number of spiral arms for model MedF is shown in Figure~\ref{fig:arms}. As expected from Equation 5, the number of spiral arms increases with radius, in this case from 2 or 3 arms close to the centre to many spiral arms at larger radii.

\section{Details of \textsc{gasoline2} simulations} 
We also ran calculations using the SPH code \textsc{gasoline2} \citep{Wadsley2017}. The main reason for using a different SPH code was that it was easier to set up initial conditions closer to equilibrium, as it is possible to run the calculations first with `particle shuffling' to settle the particles \citep{McMillan2007}, whereby particles are repositioned azimuthally to remove structure (this is difficult with \textsc{sphNG} because each $N$-body particle has a unique gravitational softening length which then becomes incorrect when particles are shuffled).  We ran a number of calculations applying the shuffling for different lengths of time, but show here a calculation where the shuffling is applied for 1 Gyr, then the simulation ran with isothermal conditions (and no star formation) for 500 Myr before cooling and star formation are turned on.

These models use initial conditions based on the \textsc{galic} code \citep{Yurin2014}, with the addition of a gas disc tailored to match the surface density of the M33 system. The stellar disc follows an exponential profile in this case, set to match the stellar surface density in the inner/mid disc. This setup also includes a live dark matter halo that extends out to nearly 500kpc. As shown in Figure~\ref{fig:velcurvesd}, the rotation curve, stellar surface density and gas surface density provide a good match to the real M33. In comparison to the \textsc{sphNG} calculations, the stellar surface density is lower, rather than higher than M33 in the outer parts of the disc. We anticipate though that the properties of the outer disc, where the surface densities are notably lower, are less critical than the inner parts for reproducing the structure of M33. 

The physics included to study the evolution in our M33 model is similar in both codes. 
For the \textsc{gasoline2} simulations we use the blastwave model of feedback \citep{Stinson2006} whereby star particles are spawned from gas particles meeting a set of criteria, which then can deliver feedback into the surrounding gas throughout the simulation. The star formation efficiency is fixed at $10\%$ in all calculations shown. We adopt varying levels of feedback efficiency, as is done for the \textsc{sphNG} calculations though in that code there is no independent star formation efficiency. All remaining parameters for cooling and star formation are the same as in \citet{Pettitt2017}. Similarly to the \textsc{sphNG} simulations, ISM cooling and heating is included. We do not consider the formation and evolution of molecular hydrogen. The stellar Toomre $Q$ and range of ISM temperatures are also similar to the \textsc{sphNG} calculations. Details of all the \textsc{gasoline2} calculations performed are listed in Table~1. 

\section{Results of sphNG calculations}
\subsection{Evolution of galaxy}
\begin{figure*}
\centerline{\includegraphics[scale=0.3]{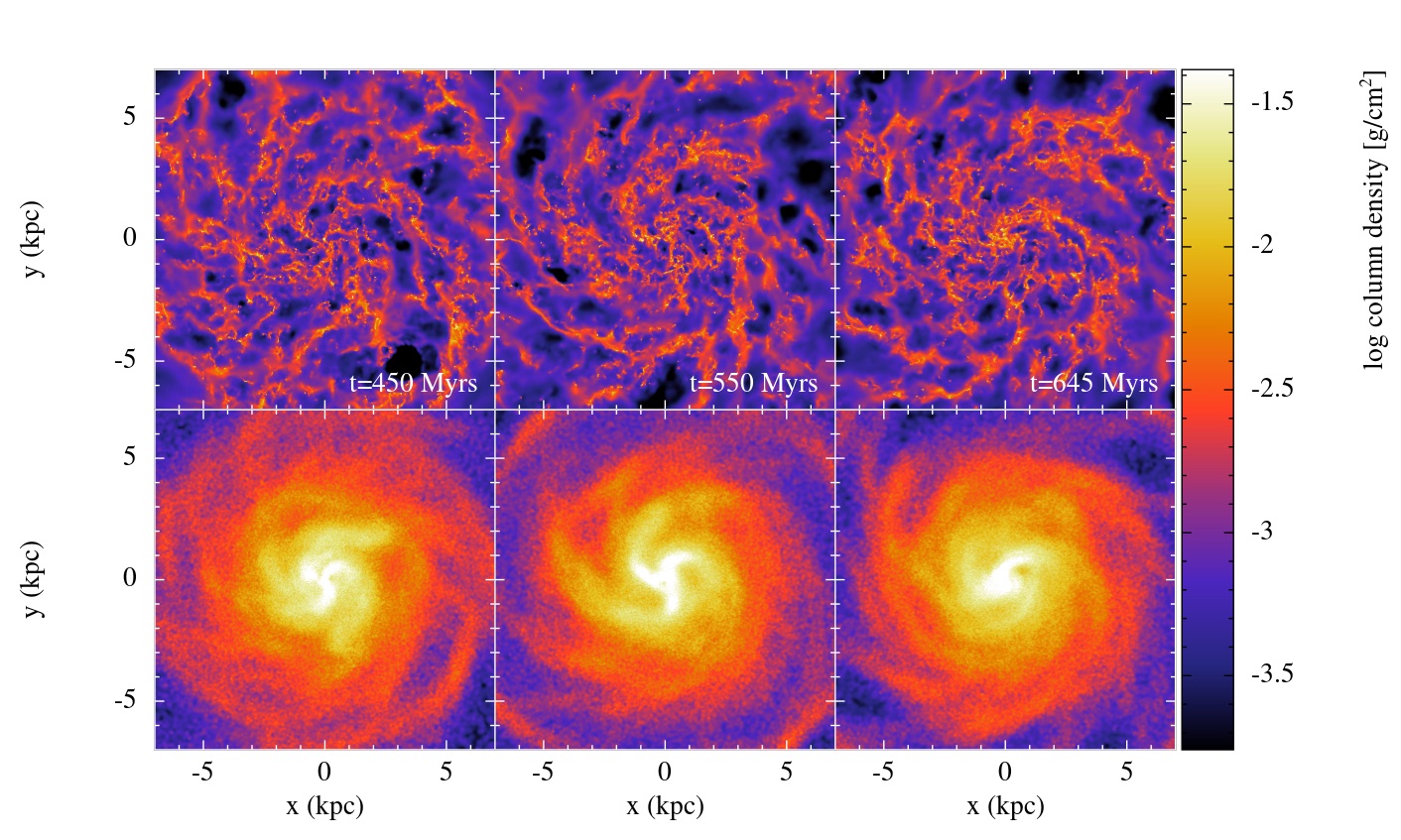}}
\caption{The time evolution of the gas (top) and stellar (lower) distributions are shown for model HighF. The times of the panels are 450 Myr (left), 550 Myr (centre), and 645 Myr (right). The gas shows a structure consisting of multiple short spiral arms. The stars have a much smoother appearance, with three more prominent arms at the centre and weak features at larger radii. }\label{fig:evolution}
\end{figure*}
\begin{figure*}
\centerline{\includegraphics[scale=0.3]{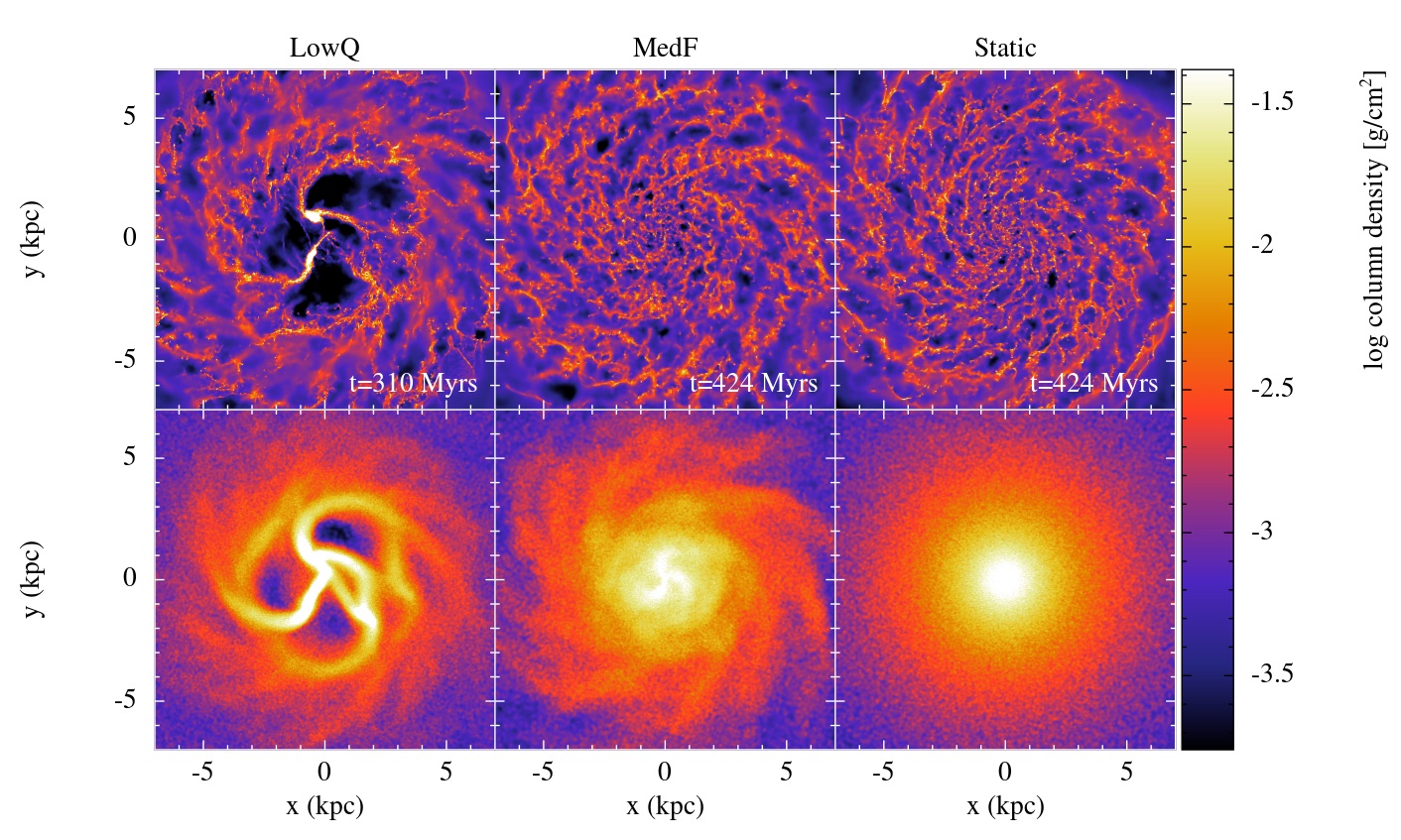}}
\caption{The gas (top) and stellar (lower) distributions are shown for models with different $Q$: LowQ (left), MedF (centre) and Static (right) with $Q\sim0.5, 1$ and $\infty$ respectively. For $Q\sim0.5$, there are very strong arms in both the gas and stars. For the static stellar disc, there are very many arms in the gas, and no clear spiral structure, as would be expected from gravitational instabilities only in the gas. For $Q\sim1$, there is slightly clearer large scale structure in the gas (this is more true for other star formation efficiencies, as we show in Figure~\ref{fig:efficiency}). The LowQ model is shown at an earlier time, as this model was so unstable it was difficult to run for long.}\label{fig:qdisc}
\end{figure*}
We show the evolution of model HighF in Figure~\ref{fig:evolution}, which is the highest feedback \textsc{sphNG} model with $Q\sim1$. We run this model up to a time of 675 Myr. The rotation period at 4 kpc is $\sim$ 275 Myr. By the times shown in Figure~\ref{fig:evolution}, i.e. a couple of rotations, any instabilities or features associated with the initial conditions tend to have largely disappeared. By 400 Myr, the evolution is fairly steady, and there is little difference between the second and third timeframes shown. The spiral structure is flocculent, with many spiral arms and arm fragments, as well as shells which are likely the result of feedback.  The stellar distribution exhibits some low mode spiral structure in the centre, and some weak spiral arms at larger radii. The stellar structure is also relatively steady with time, although slightly weaker at the last time frame. Although the structure in the gas is quite complex, there is some correlation between some of the more prominent spiral features in the gas, and the underlying structure in the stars. This suggests that the gravity of the stars is driving some large scale structure, but the processes in the ISM are also significantly contributing to the large scale structure of the gas in the galaxy.

\subsubsection{Comparison of models with different Q / static stellar disc} 
In Figure~\ref{fig:qdisc} we show models with Q$\sim0.7$ (LowQ), 1 (MedF) and our Static model. The model with $Q\sim0.7$ is least stable and produces very strong spiral arms in the stars. This causes this calculation to run very slowly, hence the time of the frame is only 310 Myr, compared to the other runs, where a time of 424 Myr is shown. For the LowQ model, the structure in the stars and gas is much stronger than the other models, and a very clear $m=3-4$ pattern is present in the centre of the disc. The increasing number of arms with radius is in agreement with Figure~\ref{fig:arms}. 

By definition there is no structure in the stars for the Static model, whilst with $Q\sim$1 there is some structure present.  Although the gas distribution for the model with $Q\sim1$ and the Static model do not appear so different, for the model with $Q\sim1$, there are gas features associated with corresponding features in the stellar density at large radii, and the large scale spiral arms tend to be a little clearer and less numerous than the Static model. For the Static model, the structure is driven by the gas physics, including self gravity, heating and cooling, and stellar feedback. This leads to more small scale structure, and multiple short arm fragments, as expected for structure driven by gravitational instabilities in the gas when the stellar disc is gravitationally stable \citep{Elmegreen1993,Elmegreen1995}. The structure of the gas in this case is a little more like for example \citet{Tasker2009}, who model a gas disc in a fixed potential with no stellar spiral structure. 

\subsubsection{Comparison of feedback efficiency}
\begin{figure*}
\centerline{\includegraphics[scale=0.3]{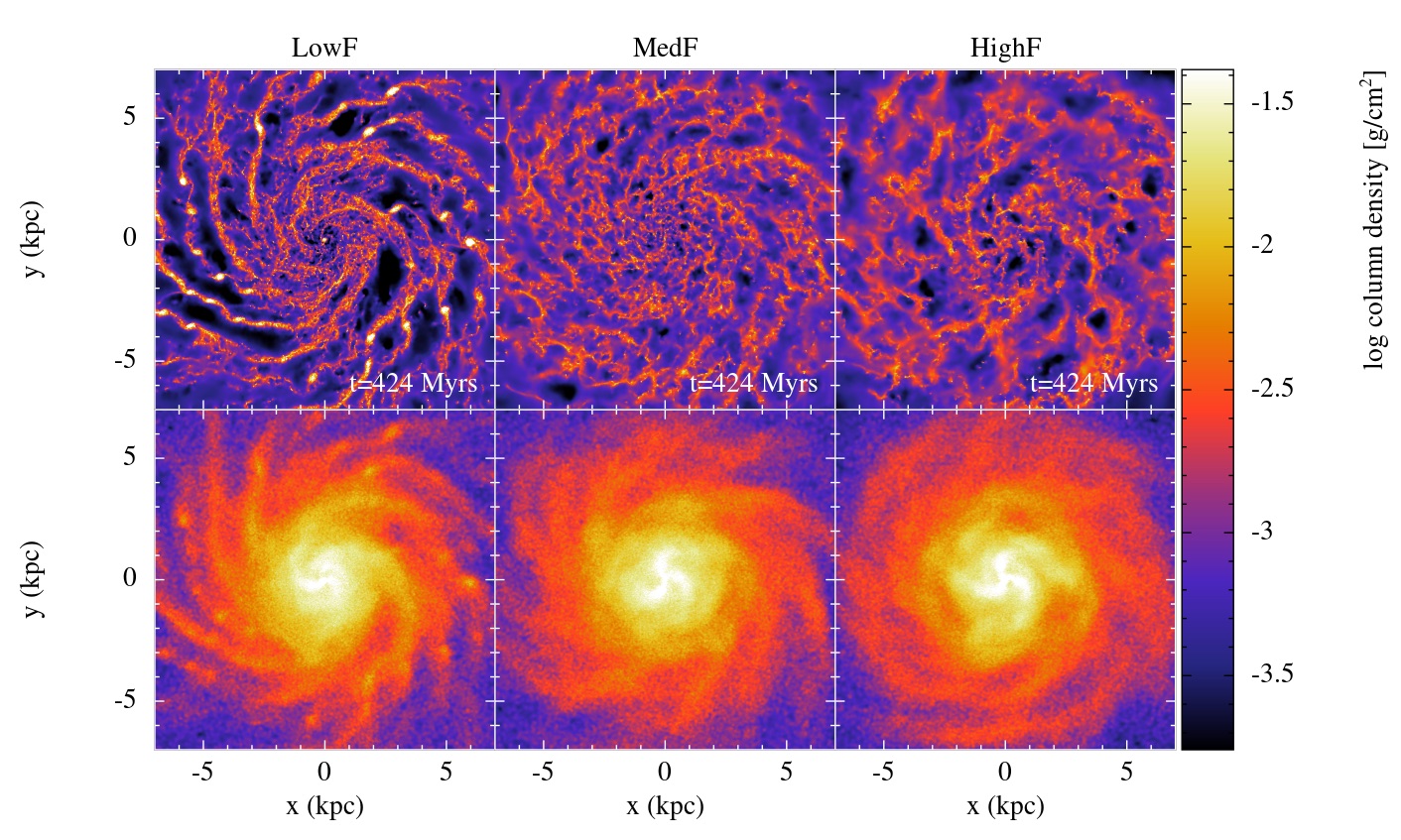}}
\caption{Models with high (LowF: left), moderate (MedF: right) and high feedback (HighF: right) efficiencies are shown. With the lower feedback efficiency the spiral arms are very clear, dense strongly bound regions of gas are clearly visible in the arms, and the stellar arms are also stronger compared to the other models. With higher feedback efficiency, the arms are disrupted by feedback, and the spiral structure is less clear in both the arms ad gas. With the highest efficiency, supernovae shells start to become clearly visible in the gas.}\label{fig:efficiency}
\end{figure*}
In Figure~\ref{fig:efficiency} we show a comparison between models with low (LowF: left), medium (MedF: centre) and high (HighF: right) star formation efficiencies. There is a considerable difference in the structure of the gas for the model LowF, with lower feedback compared to the other simulations. This model produces longer, clearer spiral arms in both the stars and gas. With low feedback, the spiral arms in the gas appear continuous.  In the other models, feedback acts to break up the spiral arms. Massive clumps of gas can also be seen along the spiral arms, which are produced when stellar feedback is unable to disrupt giant molecular clouds. For the low efficiency case, there is also a very clear correspondence between the gas and stars. The stellar arms are also more pronounced than any of the other models. Thus in this model, the gas is evidently contributing to the generation and maintenance of stronger spiral arms in the stars.  

The models with the higher efficiencies both show similar, relatively weak structures in the stars, and less clear arms in the gas.  For the highest star formation efficiency model, there actually appear to be slightly less numerous, more continuous spiral arms than the MedF model. This is presumably because the higher feedback disrupts, or even prevents the formation of weak features in the gas. The higher feedback is also able to blow out holes and shells, which are not evident in the other models. The asymmetry of the shells, and their aspect ratios compared to LowF,  distinguish the low density regions in model HighF as shells, rather than simply low density regions between the spiral arms.

\subsection{Explanation of models}
Our models show that both the underlying gravity of the stellar disc, and the properties of the gas, determine the large scale spiral structure in the stars and gas. The spiral arms in the stars are formed by gravitational instabilities in the disc, primarily in the stellar component. We hypothesise here that the spiral arms are primarily driven by gravitational instabilities in the stars, but that gas allows dissipation of energy in spiral shocks, which helps maintain the spiral arms. As discussed by \citet{Kalnajs1972},  in the absence of gas, as the stellar density increases locally, the stellar arms cannot dissipate energy and instead the velocity dispersion increases. When gas is present though, energy can instead be dissipated by the gas in spiral shocks. The amount of energy dissipated will depend on the compression of the gas. We test this hypothesis by comparing the evolution of galaxy models which only contain stars, and comparing $Q$ for our models. 

We note that in the case of transient spiral arms, the difference in pattern speed between the spiral arms and the angular velocity of the galaxy at a particular radius may be relatively small, particularly compared to grand-design type spirals \citep{Wada2011,Grand2012}. However when gas is cold ($<$ 1000 K), the sound speed becomes less than 1 km s$^{-1}$, and the gas can still experience a shock with a relatively small difference in velocities.

\subsubsection{Evolution of  $Q$}
\begin{figure}
\centerline{\includegraphics[scale=0.4]{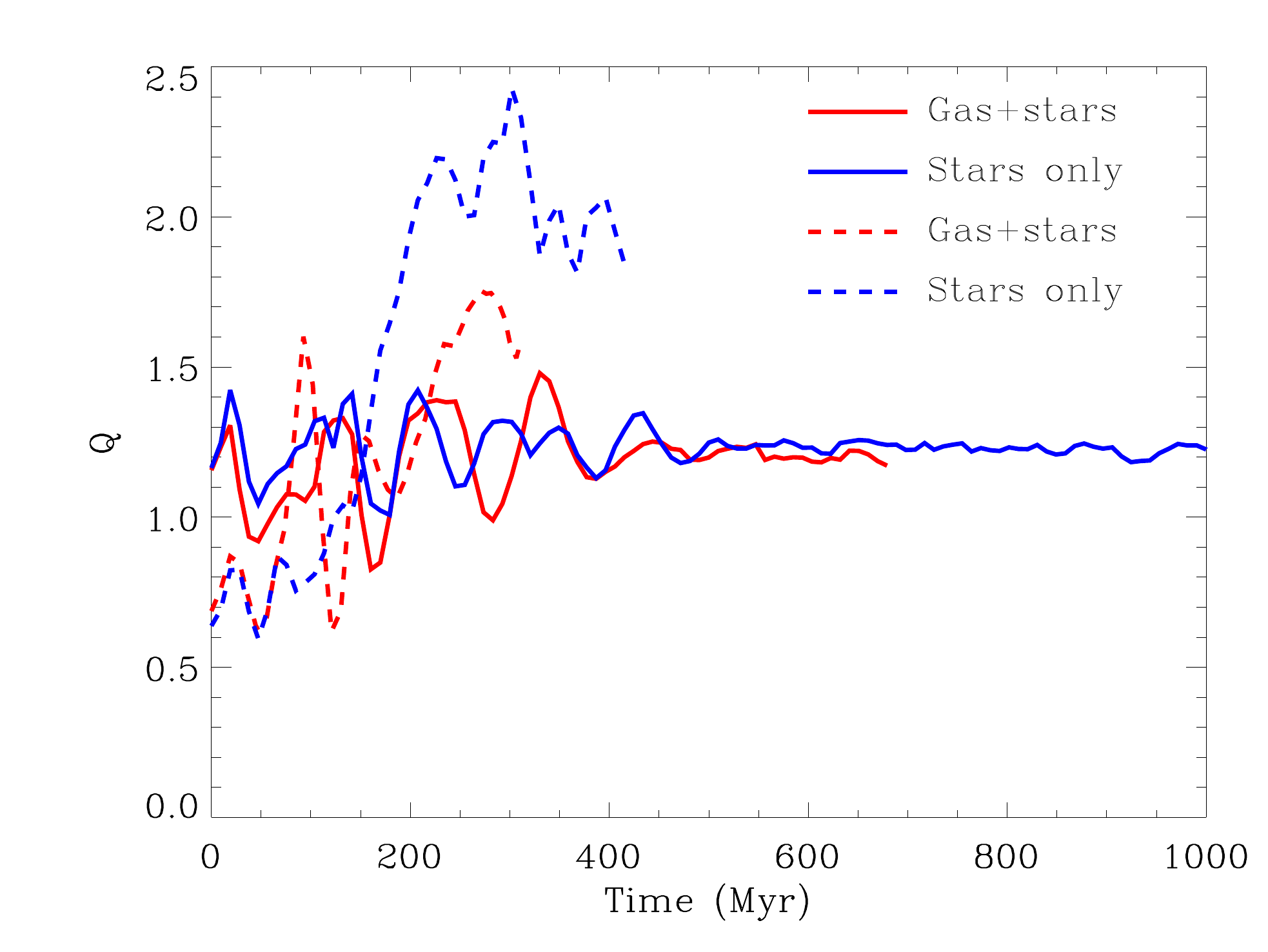}}
\caption{$Q$ for the stellar component is plotted for the models MedF (blue solid line), StarsOnly (red dashed line), LowQ (blue dotted line) and StarsOnlyLowQ (red dashed line). These models show the change in Q with (red) and without (blue) gas, starting with low (dashed) and higher (solid) values of $Q$. When the models start with $Q \gtrsim 1$ there is little evolution in $Q$. However if the disc is initially unstable, $Q$ increases significantly, more when there is no gas compared to without. }\label{fig:qplot}
\end{figure}

In all of our models, $Q$ for the stars is lower than $Q_{gas}$, typically by a factor of several. This indicates that self gravity will be significantly stronger in the stars than gas, and the stars, rather than gas, are likely to drive large scale structure. To examine the role of gas in producing and / or maintaining the spiral arms, we also run a couple of models with no gas (StarsOnly and StarsOnlyLowQ). The difference between these models is that $Q\sim1$ in StarsOnly and $\sim0.7$ in StarsOnlyLowQ. We show in Figure~\ref{fig:qplot}  the evolution of Q in these models, and corresponding models with gas, MedF, and LowQ. For all models, $Q$ is calculated at a radius of 4 kpc. Figure~\ref{fig:qplot} indicates that for the models with a larger initial $Q$, there is no increase in $Q$, for both models with and without gas. However for the models which start highly unstable, with $Q\sim0.7$, there is a large increase in $Q$. This is consistent with the work of \citet{Fujii2011}. Provided the resolution is sufficient, these models with higher $Q$ only show an increase in $Q$ over timescales of several Gyr, much longer than we can reasonably simulate with our models with gas. Presumably these models form weaker spiral arms, the velocity dispersion of the stars remains fairly low, and  the disc heats up only very slowly. The models with low initial $Q$ in \citet{Fujii2011} also show a large increase in $Q$ over short timescales. In this case, the disc forms very strong arms, which presumably heat the stars up much quicker, and consequently the spiral arms in these unstable discs (and similarly spiral arms in under-resolved discs) are very short-lived. Unfortunately because the disc is so unstable, it is difficult to run our model with gas very long, but we nevertheless see a smaller increase in $Q$ for our model with gas compared to with only stars. For the models with low $Q$, the gas does not make so much difference, because there is little change in $Q$, but our work suggests that gas may have a greater impact on more unstable models. Consequently, this suggests that to fully test the role of gas, and the ideas presented in \citet{Kalnajs1972}, we would need to consider relatively unstable gas discs. 

We also tested how $Q$ varied between the models at other radii. At smaller radii, we find similar behaviour, but at large radii, the change in $Q$ is fairly small for all models, and we don't see a difference in behaviour with or without gas.

We can also compare the models with different levels of feedback. We would expect that in the models with low feedback, because there is more cold gas, the spiral shocks are likely to be stronger and dissipate more energy. Visually, a comparison of the spiral arms for the two models LowF and HighF matches this expectation. The stellar spiral arms are clearly stronger when there is a low level of feedback, particularly at larger radii. We might expect $Q_{stars}$ to be lower for the low feedback case, however we find that $Q$ is the same, both for the models with different feedback and different temperature thresholds. It is not clear why there is so little difference in Q, but it may be that gas has a non-negligible role in determining the stability of the disc.  As mentioned in Section 2, it is difficult to constrain the combined gravitational instability of the gas and stars to a single parameter, because the gas is a multiphase medium. However in LowF (the low feedback case), more of the gas is cold (few 100 K) compared to the other models. Thus the gas is more unstable, and if the sound speed for the cold gas is used in a calculation of $Q$, we find that the $Q$ for the gas is only a little higher than $Q$ for the stars. In other models, where the gas is predominantly warm ISM, using the sound speed to calculate $Q$ for the gas still gives a value at least several times higher than $Q$ for the stars.

\begin{figure}
\centerline{\includegraphics[bb=0 170 600 650, scale=0.36]{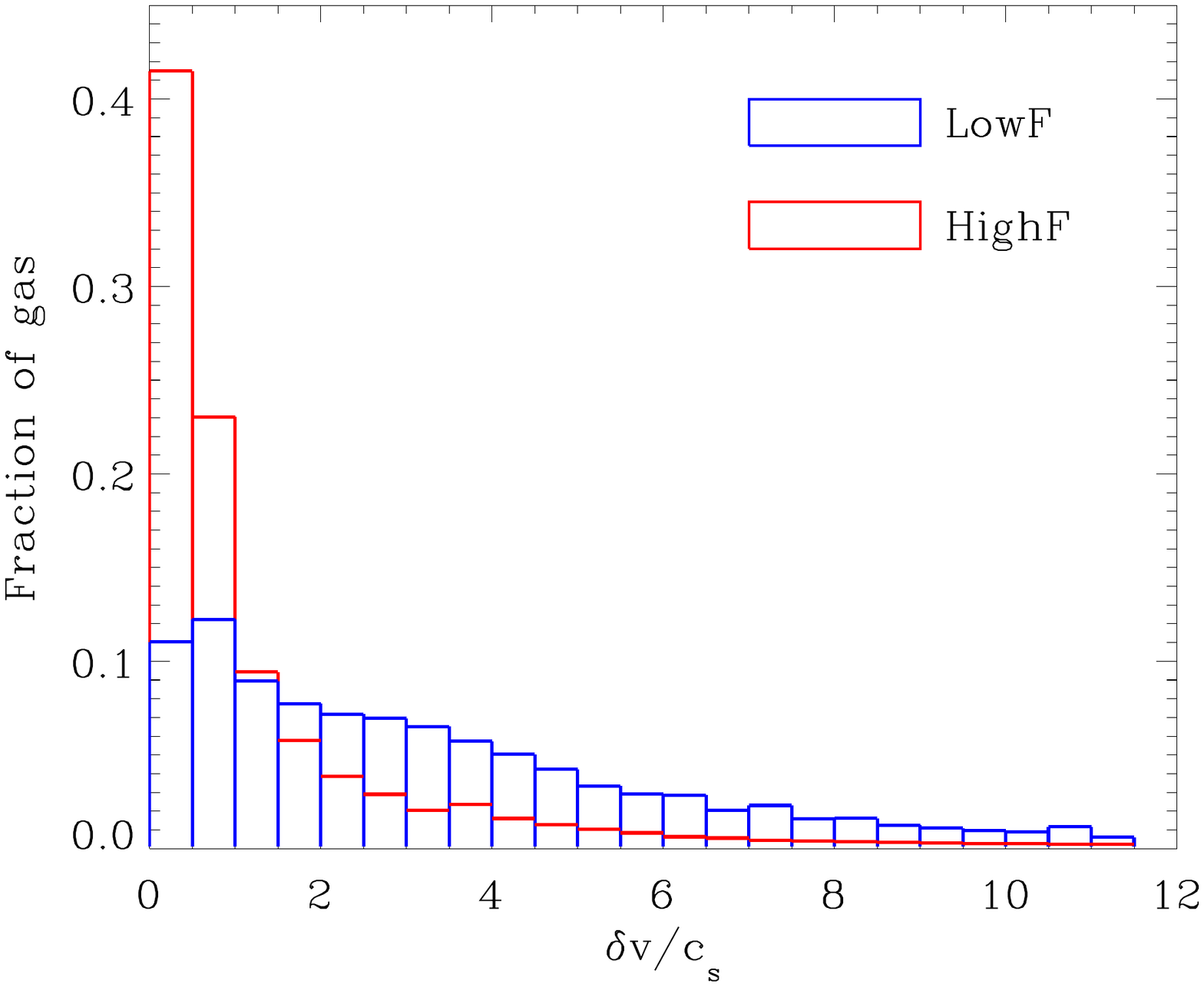}}
\caption{Fraction of gas exhibiting different values of $\delta v/c_s$ where $\delta v$ represents the change in velocity of the gas as it crosses a spiral arm, and thus $\delta v/c_s>1$ represents shocks in the gas. The model with the low feedback efficiency exhibits a greater fraction of gas undergoing shocks, and thus the gas produces a stronger response to the underlying stellar structure, and clear spiral arms are seen. The model with the high feedback has less gas with large $\delta v/c_s$, and thus does not have such a strong response to the stellar structure.}\label{fig:velocities}
\end{figure}

\subsubsection{Response of the gas}
Whether the gas shocks, and how strongly the gas shocks in response to stellar perturbations will depend on a number of factors including the strength of the perturbations, their pattern speed relative to the rotation of the galaxy and the sound speed of the gas. \citet{Binney2008} present a simple toy model for the response of gas to a potential, whereby a `force factor' determines whether the gas undergoes a shock. They assume a fixed potential, but we adopt a similar idea with an evolving potential. We examine a similar type of condition, but instead use the change in velocity, $\delta v$ due to the stellar potential and the sound speed of the gas.  If $\delta v/c_s>1$ then we would expect the gas to shock. Because the stellar potential varies with radius, we determine $\delta v$ in radial bins of width 0.5 kpc. We compute $\delta v$ by averaging the maximum change in azimuthal velocity for all the gas particles within a given radial bin, over a time period of 200 Myr. For each particle, we then compute $\delta v/c_s$ using the $\delta v$ averaged for the radial bin where the particle is located, and the sound speed of that particular gas particle.

We show histograms of $\delta v/c_s$ in Figure~\ref{fig:velocities} for the models with high  (HighF) and low feedback efficiency (LowF). Figure~\ref{fig:velocities} indicates that the LowF model has a higher amount of gas with low $\delta v/c_s$, whereas the HighF model tends to exhibit lower $\delta v/c_s$. This again suggests that gas in the model with lower feedback is better able to shock, producing both more pronounced features in the gas and lowering the total $Q$, resulting in a clear multi-armed spiral structure. In contrast, for the model with higher feedback the gas tends to be warmer, so the gas has less ability to shock. Consequently features in the gas are less pronounced and a more flocculent structure is apparent. 

\subsection{Comparison with actual M33}
\subsubsection{HI}
\begin{figure*}
\centerline{\includegraphics[scale=0.6]{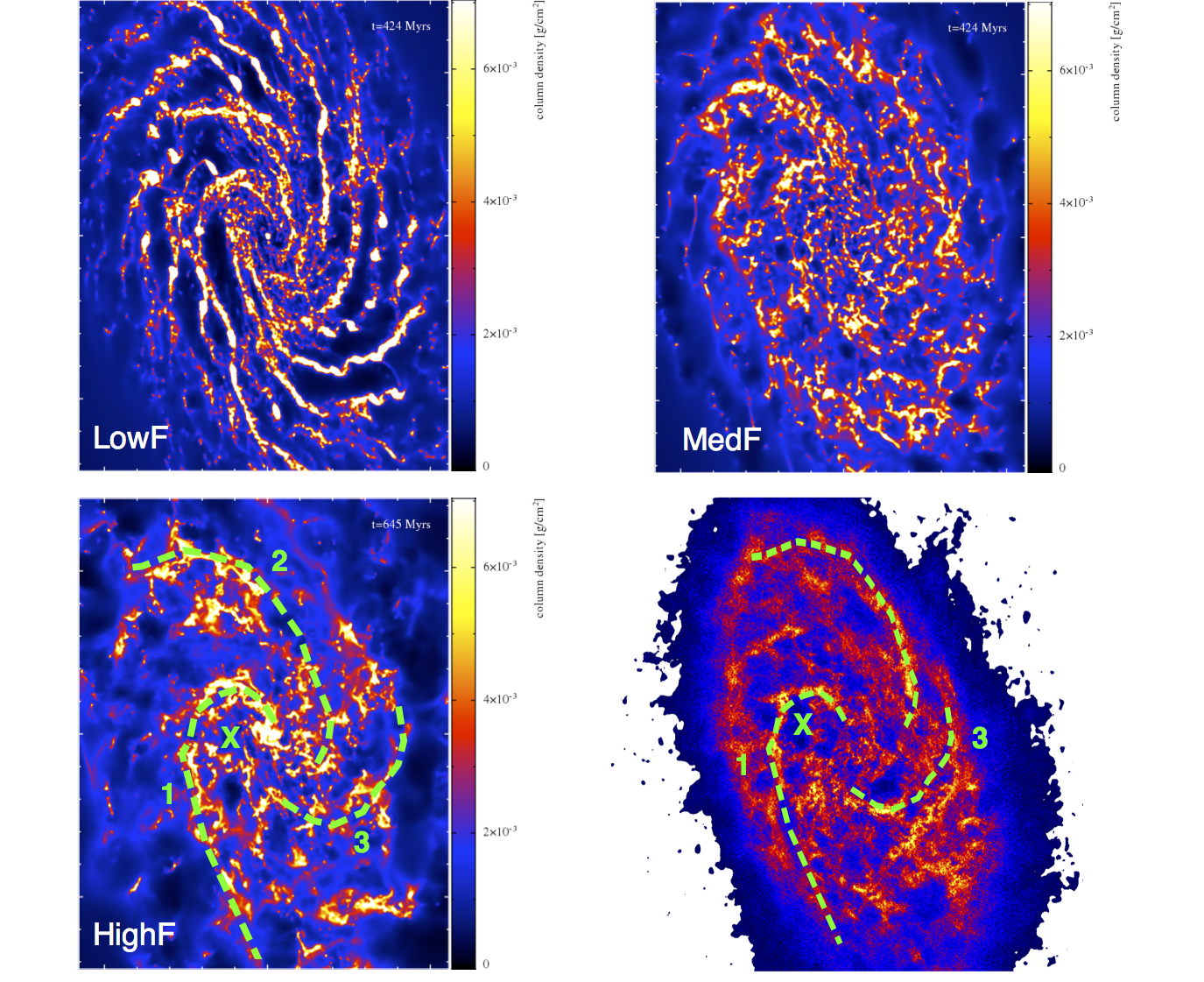}}
\caption{Column density images for different models presented in this paper compared with the actual M33 HI emission. All scales are linear, both for the observations and simulations. The top an bottom left panels show the models with different levels of feedback. The low feedback model (LowF) shows very strong spiral arms, and very bright dense clumps which are not seen in the actual M33. The medium feedback model shows less clear spiral arms compared to the actual M33. The best match by eye is the high feedback case. Here the feedback is sufficient to create large holes or shells, which also appear to shape the gaseous arms and push the gas into a smaller number of arm features compare to the models with weaker feedback. For the models MedF and HighF, the timeframes are shown when there is best agreement by eye with the real M33. Of all the models, only the HighF model shows particular differences in the structure (namely the appearance of large shells) with time such that the time of the snapshot is significant.}\label{fig:observations}
\end{figure*}

In Figure~\ref{fig:observations} we show column density plots from some of the simulations compared to the HI map of \citet{Corbelli2014}. The observations of M33 were made with the VLA and GBT and achieved a spatial resolution of 10'', or 41 pc at the distance of M33 (taken to be 840 kpc).  
The simulated galaxies have been rotated to match the orientation of M33. We take an inclination angle of 54$^{\circ}$ and position angle of 23$^{\circ}$, from \citet{deVauc1991}. Note the simulated galaxies have also been reflected so that the spiral arms rotate in the same direction as M33. Only the model with high feedback, HighF, shows a particularly time dependent morphology. For this model, we chose a timeframe where the simulated galaxy best resembled the actual M33, whilst the other models are shown at timeframes of around 425 Myr. The simulated galaxies tend to show sharper and brighter features compared to the real M33, whereas the HI map for M33 appears much smoother (this is slightly less so for the model with the higher temperature threshold, where presumably the gas is smoothed out more). One possible explanation for part of this difference is that in reality the denser features may be molecular, and so not appear in the VLA$+$GBT map. Previous studies show that for a whole galaxy, it is difficult to resolve the molecular component - only by modelling subregions of a galaxy or individual clouds are realistic molecular gas densities achieved \citep{Duarte2016,Rey-Raposo2017}. A second factor is that the resolution of the simulations is $\lesssim10$ pc in the denser regions, which is a factor of 4 or more higher than the M33 HI map.  

The model with the highest feedback, HighF appears to show the best agreement with the real M33, in terms of spiral structure in the gas. Both the simulated and real galaxy appear to contain 3 dominant spiral arms and some fragments of spiral arms. We have highlighted the three main spiral arms in the HighF model with dashed lines. These lines were produced by dividing the galaxy into rings and finding local density peaks in the density versus azimuth. Density peaks at different radii are then assigned as a spiral arm or section. The spiral arm labelled `2' comprises two sections joined together, whilst arms 1 and 3 are each one section. The lines were determined for a face on map and then rotated. For the image of the real M33, arms 1 and 3 have simply been copied and pasted from those shown for the HighF model, indicating that the simulations have reproduced the shape snd position of the arms remarkably well. Arm 2 has also been copied and pasted but shifted relative to arms 1 and 3. The shape and position of the arms are extremely similar in both the real and simulated galaxy, though in the actual M33 there is a stronger arm feature slightly lower than the plotted arm 3 (this arm in the actual M33 appears to consist of two sections, one of which is coincident with the plotted arm 3 and one which is just below this). Interestingly arm 2 in the simulation arises partly from chance alignment of sections which are not connected when viewed face on. As well as the labelled features, there are also additional fragments of spiral arm below arms labelled 2 and 3 in both the simulation and actual M33.

 As mentioned above, we selected this timeframe of the HighF model as that when the simulated M33 matched the real M33 particularly well, predominantly focusing on the shape and location of arm 1. We actually find two good matches between 400 and 700 Myr, at the time shown (645 Myr) and at 440 Myr. These timeframes require the spiral arm features both to match the actual M33 features, and lie in the correct orientation, In particular these times are characterised by the ability of stellar feedback to produce a shell, or multiple shells which shape arm 1 and lead to a clear spiral arm  at this location in those time frames. The low density regions marked with an `X' are shells blown out by the supernovae feedback shaping the spiral arm, and correspond to a large shell in M33 associated with NGC604 (again marked with a `X'). Large shells produced by feedback are a characteristic of the HighF model, which are not readily apparent in the other models, but again do match features seen in the actual M33. We note that the spiral arms in the gas are quite short-lived as they are readily dispersed by feedback, and only last 10s of Myrs.
 
In contrast to the high feedback model, the models with low efficiency, LowF and MedF show worse agreement. For model MedF there is little large scale structure in the gas, thus we can conclude that a model that doesn't produce some large scale structure will not well represent M33. By contrast, the model with the low feedback efficiency shows too much large scale spiral structure, and the spiral arms appear to be more coherent than the real M33. Thus overall, our models seem to suggest that the structure of M33 is due to a combination of gravity of the stars and gas, producing an underlying spiral perturbation, and stellar feedback which disrupts the spiral pattern, sometimes even reducing the number of spiral arms present in the gas, and creating shells which again may replace spiral structure.  
 
\subsubsection{Stars}
\begin{figure}
\centerline{\includegraphics[scale=0.3, angle=90]{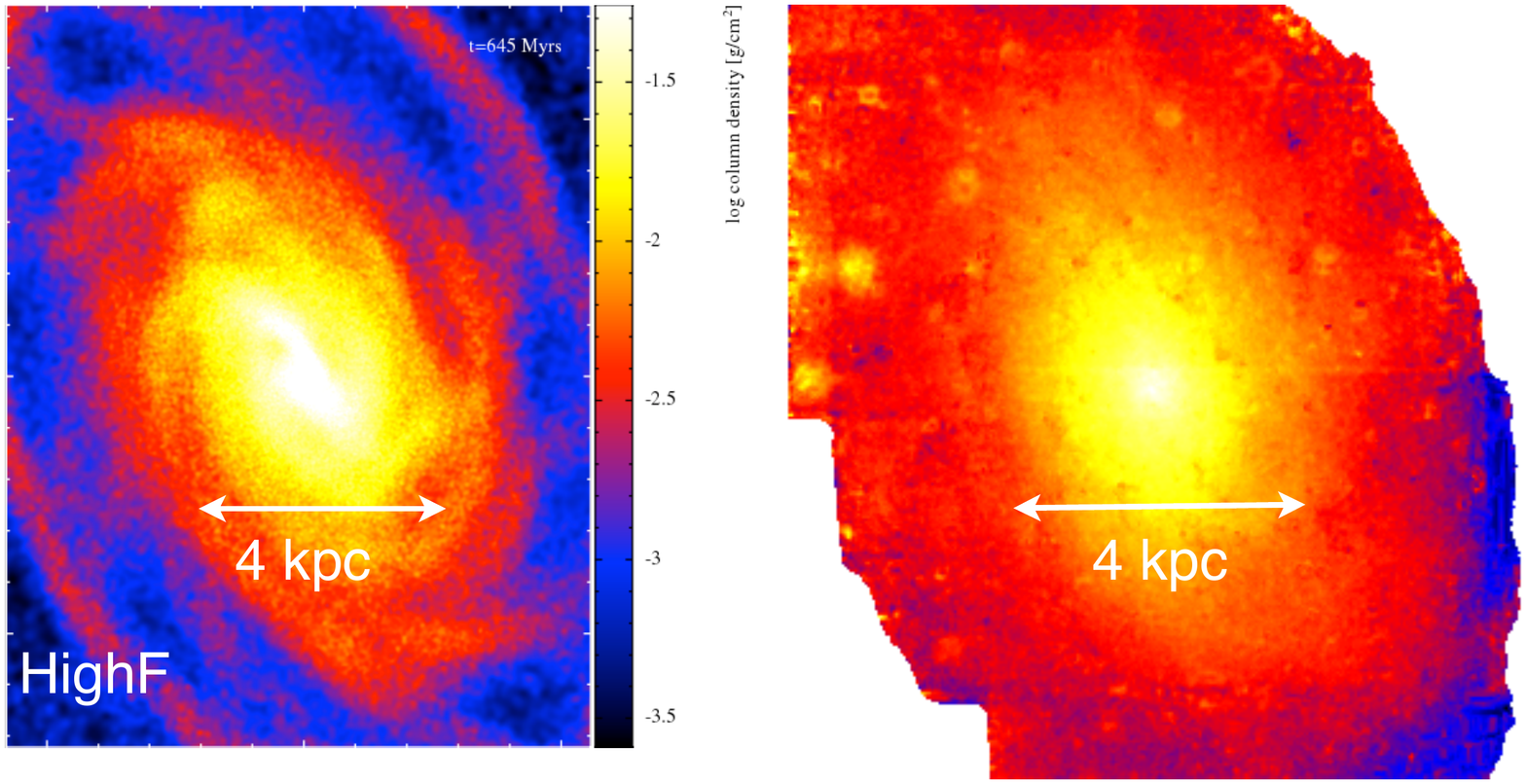}}
\caption{The stellar density map is shown for the model HighF (left), from a time of 645 Myr, and the actual M33 (right). Both show a logarithmic scale, but otherwise the scales are not chosen to match. Both the model and the real M33 show very little spiral structure.}\label{fig:observations_stars}
\end{figure}

Figure~\ref{fig:observations_stars} compares the stellar density from the HighF model (left) and M33 (right) from \citet{Corbelli2014}. Both figures use a logarithmic scale, although otherwise the scales are not chosen to match. The stellar density map of M33 is derived from a comparison of the synthetic spectral energy distribution with multi-band optical imaging and reaches the sensitivity limit in the outer disc (blue color in Figure~\ref{fig:observations_stars}.  Both the model and M33 show very weak spiral arms. For the observations, it is difficult to distinguish any spiral arms, although by changing to a histogram scale with ds9 it is possible to see some spiral structures. For the models we note that the strength of the spiral arms varies a little with feedback, and with time, as they get slightly weaker in the HighF model at the latest times of $\gtrsim$ 675 Myr (so earlier timeframes in the models show worse agreement with the observations).

\subsection{Higher resolution models}
\begin{figure*}
\centerline{\includegraphics[scale=0.3]{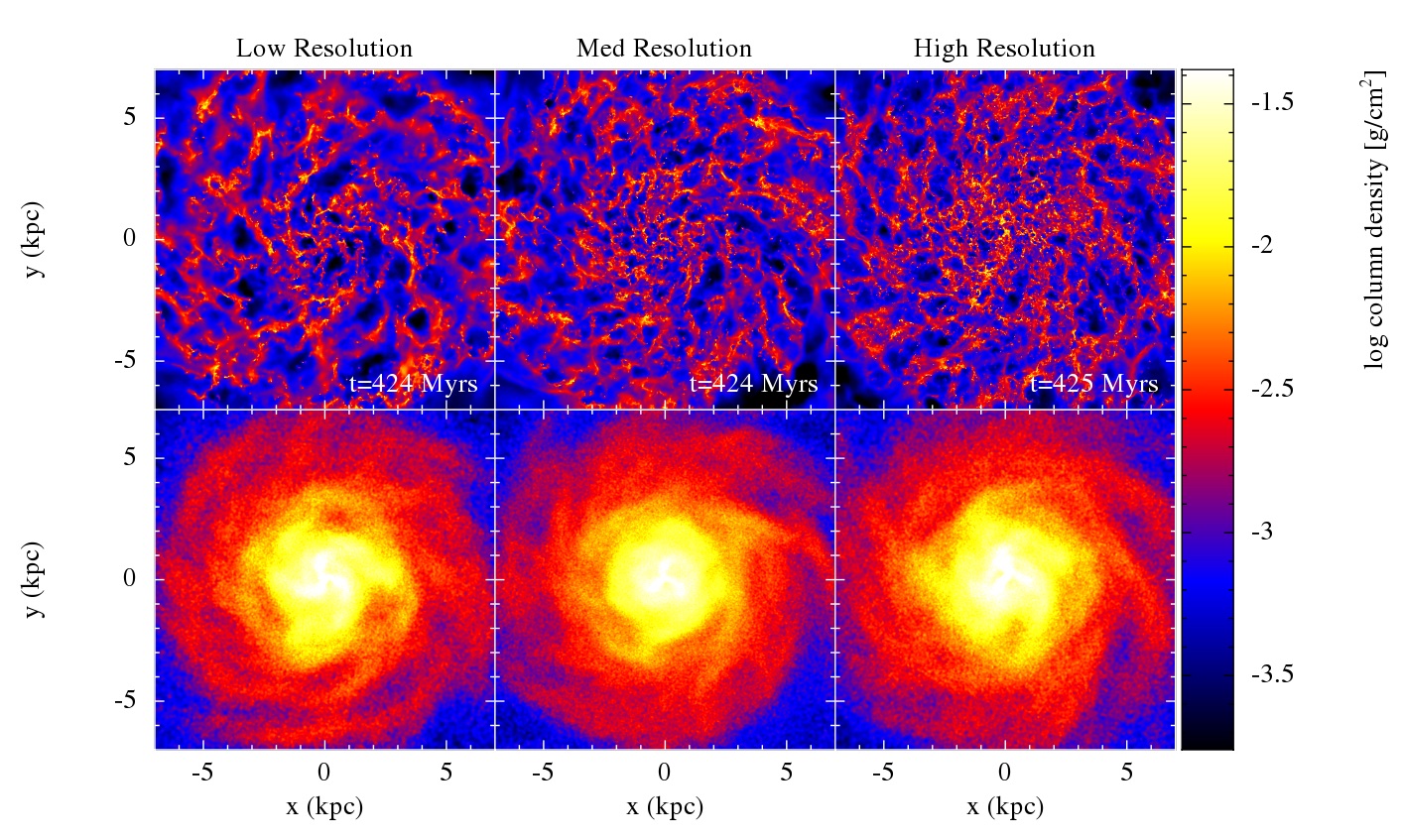}}
\caption{The gas (top) and stellar (lower) surface densities are shown for different resolution \textsc{sphNG} models at times of 424 Myr (HighF, MedRes and HighRes). The number of gas particles is 137500 (left), 2750000 (centre), 7087500 (right). The structure of the gas looks fairly similar at different resolutions, in particular the effects of feedback in driving the gas structure. The feedback appears slightly less effective at higher resolution. There are some differences in the stellar disc, but these are attributable to small differences in $Q$ for the different models.}  \label{fig:res}
\end{figure*}

We compare sphNG models with the standard (HighF model), and two higher resolutions, MedRes and HighRes in Figure~\ref{fig:res}. All the models use the same feedback efficiency. The models appear relatively similar in both the gas and stars. In fact the highest resolution model is more similar to the low resolution model than the medium resolution case. There are some differences between the simulations, at higher resolution there is slightly more substructure in the gas, whilst in the medium resolution model the spiral structure in the stars is slightly less clear. We attribute these differences to the difference in behaviour of the feedback in the different models, and the stability of the disc, both of which are difficult to exactly replicate when changing resolution. The holes are perhaps not quite so clear in the higher resolution models, likely because the same amount of feedback is not quite so effective at the higher resolution. There are also differences in the evolution of $Q$ for the models (at the few \% level), even though $Q$ is initially 1, which likely explain the small differences in the stellar disc at the different resolutions. Generally we found that quite small differences in $Q$ lead to somewhat different stellar disc features, perhaps suggesting that our models could be further fine tuned to better represent the stellar structure of M33 if needed. Although not shown, a model with a lower feedback efficiency was also performed at the same resolution as model HighRes. This tended to show quite similar structures to the corresponding low resolution model. For the lower feedback case clearer spiral arms are not seen either at low or high resolution. 

\section{Results of GASOLINE calculations}
\subsection{Global properties}
We now turn to our second set of simulations, those made with the cosmologically-focused SPH code \textsc{gasoline2}. While initial conditions are in general agreement there are a number of subtle differences between these and the simulations already discussed (both inherent to the code and in the initial conditions). The main difference is that these models are in much better equilibrium at initialization compared to the previous models. 

\begin{figure*}
\centerline{\includegraphics[scale=0.3]{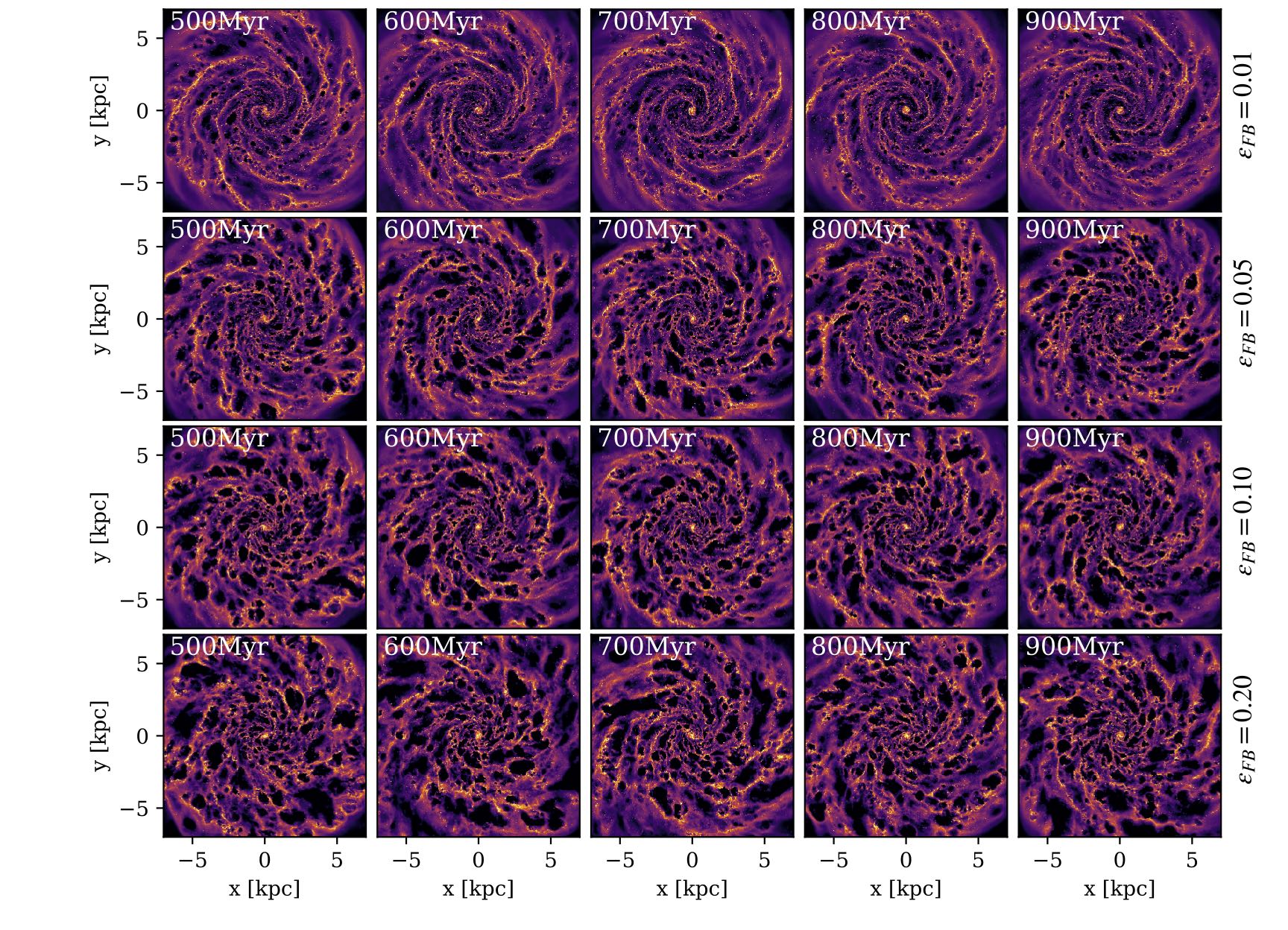}}
\caption{Time series of gas in all \textsc{gasoline2} calculations. Each column shows a different time, and each row a different level of feedback.}\label{fig:gasolinetime}
\end{figure*}

Figure\;\ref{fig:gasolinetime} shows the evolution of the gas structure in all four gasoline calculations, with the only change being the feedback efficiency level. Each row shows a different model at 5 different time-frames. While the discs are in very good dynamical equilibrium at $t=0$Myr, as soon as feedback occurs a shockwave radially propagates through the gas disc, caused by the initial burst in star formation as the gas cools. This resulted in a ring-like feature in the 20\% model, and a very weak ring in the 10\% model, but was not visible in the weaker feedback cases. However, this ring rapidly disintegrates after 300--400Myr, and the systems then enter a more stable evolution phase. As such, we only show results after 500Myr of the activation of star formation and feedback.

In Figure\;\ref{fig:gasolinetime} a clear correlation can be seen between the strength of feedback and the structure of the ISM gas. The weakest feedback model has very well defined spiral arms and a smooth inter-arm appearance. These dynamic spirals come-and-go as the simulation progresses, but the amount of structure in the disc is well maintained throughout the entire 500Myr shown. The remaining models have a very similar structure, with only small segments maintaining an arm-like structure. The size of cavities in the gas increases with feedback efficiency, as expected, with the strongest feedback model creating cavities greater than 2kpc in diameter. For the 10\% and 5\% model these cavities seem to reside in mostly the outer disc, where the lower stellar surface density and higher levels of differential rotation at larger radii allows for supernova driven cavities to shear out into larger voids without filling up with gas as fast as those in the mid/inner disc. The 20\% model has strong enough feedback to excavate large cavities at smaller radii.

The overall behaviour of the \textsc{gasoline2} models is roughly similar to the \textsc{sphNG} models, in particular the finding that with a low feedback efficiency, the arms are too continuous, and the large holes visible in M33 are not well produced, whereas better agreement is found with higher levels of feedback. The models with higher levels of feedback look fairly similar to the MedF model from the \textsc{sphNG} simulations. The HighF \textsc{sphNG} model tends to have somewhat larger voids, and spaces between the arms than the \textsc{gasoline2} models.

\subsection{Observational diagnostics}
\subsubsection{Gas}
We now present numerous tests of the models compared to various different observational constraints, as was done with the \textsc{sphNG} models. We also make some additional comparisons using software designed for comparing \textsc{gasoline2} models to observations.

\begin{figure*}
\centerline{\includegraphics[scale=0.9]{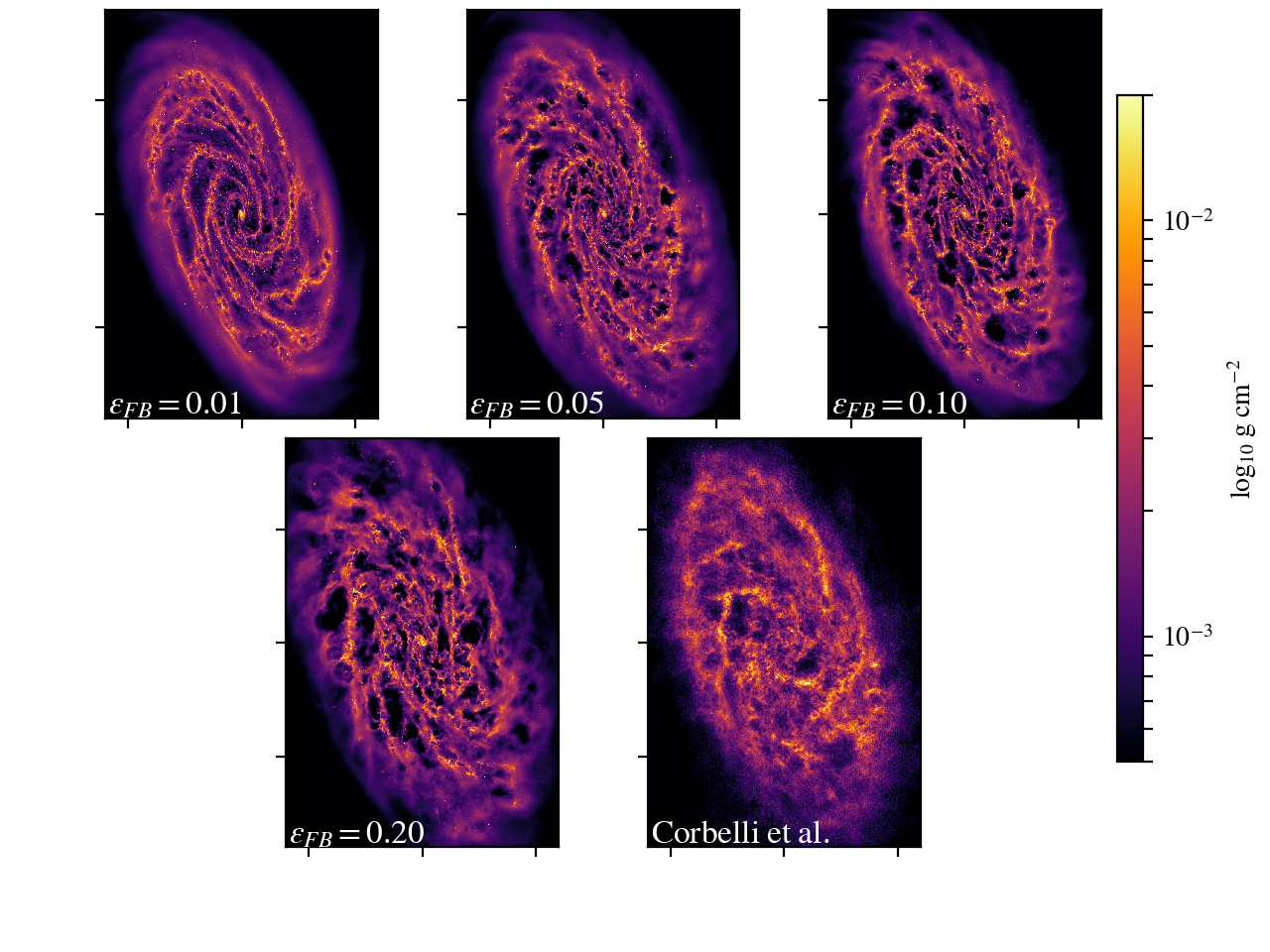}}
\caption{Best fitting \textsc{gasoline2} models compared to M33 gas distribution (lower right).}\label{fig:gasolineside}
\end{figure*}

Fig\;\ref{fig:gasolineside} shows a time-frame of each of the \textsc{gasoline2} simulations compared to the gas surface density data of \citet{Corbelli2014} in the bottom right. Best fits were found via a combination of by-eye and an automated routine using a structural similarity index (SSIM, built into \textsc{python}'s \textsc{scipy} module) of images of the gas surface density. The simulated disc is first inclined at the orientation of M33, then rotated in increments to assess the suitability of different phase angles. These images are smoothed over a scale of 2kpc to leave only the strongest features. The SSIM is then calculated for each time-stamp and phase angle ($\Delta t=10$My, $\Delta\theta=10^\circ$ using the time domain of 0.5--1.0Gyr, resulting in 1800 images for each calculation). The best fitting of these low resolution comparisons are then shown in Fig\;\ref{fig:gasolineside}, now at their natural resolution.

In Fig\;\ref{fig:gasolineside} we show one of the better-fitting snapshots of each simulation, orientated in the same manner as the observed M33. Certain features like the arms shown in Fig\;\ref{fig:observations} are seen in some of these models, with accompanying inter-arm voids. The lowest feedback model displays some very similar features; the upper-right arm structure, the irregularly shaped arm north-west of the centre, and the straight-arm segment in the edge of disc to the south-east. The nature of dynamic spiral features such as these allows for spiral arms that are less regular than the assumed fixed log-spiral features used by density wave-like potentials \citep{Grand2012,Baba2013}. Cavities in the gas seen in M33 are a simple result of interarm voids. However, the gas does not display the same smaller scale features, like the irregular patchy features seen in the inter-arms of M33.

The 5\% feedback model shows a more disrupted spiral structure. Similar small arm features are seen as previously, though now the increased feedback stunts their growth considerably. This appears one of the poorer matches to the observational data, with both the feedback cavities and inter-arm regions too small to provide a good match for M33.

The remaining higher feedback models provide a good match to the general structure of M33, though neither is a precise match. The 20\% model is very effective at creating large ISM cavities, much like the \textsc{sphNG} models, though this comes at the expense of creating very few strong arm features. The cavities in the 20\% model may even be too large and numerous, with the disc perhaps showing too great a degree of fragmentation compared to M33. The 10\% feedback model is a good middle ground, creating a handful of large cavities but also managing to maintain some elongated arm features.

We found that many of the snapshots not shown here produced certain features very well, while missing others entirely. Enough realizations could feasibly eventually produce every feature simultaneously, though such a brute-force approach is hardly practical. It may also be that our choice of $Q=1$ is somewhat too stable to form the features seen. A higher $Q$ value will promote the growth of features somewhat, overcoming some of the disruptive effects of the higher feedback and producing the higher arm-interarm contrast seen in the M33 gas. Though this brings with it accompanying problems, most notably in increasing susceptibility to bar formation. A small bulge component has already been added to this model to suppress bar formation, and decreases in $Q$ will only accelerate this instability further.

All discs show a bright inner spot, which is not seen in most of the \textsc{sphNG} models, though there is a slight inner concentration in the HighF model at later times (Fig.\,\ref{fig:evolution}). We believe this is due to the live spherical mass distributions now present in the \textsc{gasoline2} models, namely a small bulge and the live dark matter halo. Momentum exchange between these features and the centre of the gas disc may cause an element of orbital decay in the gas, resulting in the build-up of small scale structure seen in these calculations. However, this high density inner gas deposit may be analogous to the peak seen in molecular gas seen in the centre of the M33 disc \citep{Heyer2004}.

\subsubsection{Stars}

\begin{figure}
\centerline{\includegraphics[scale=0.5]{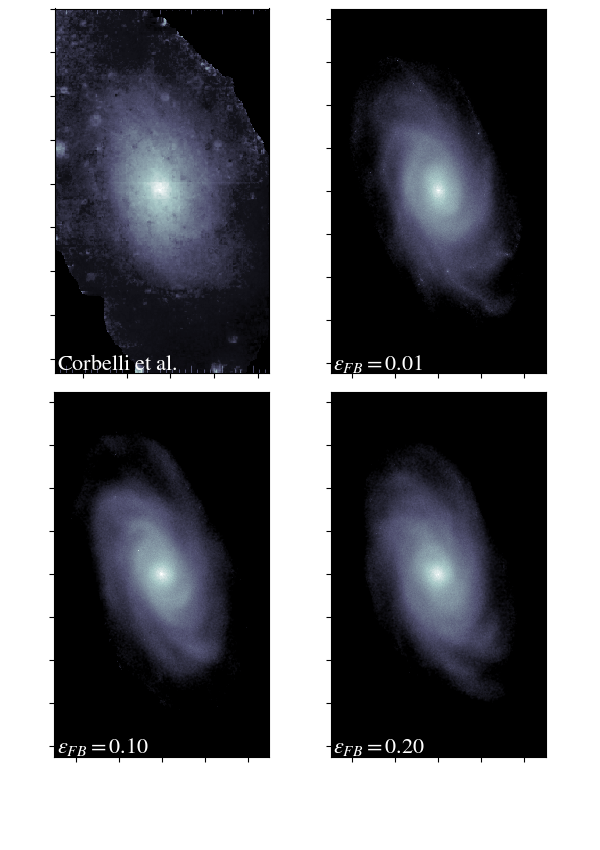}}
\caption{Stellar surface density maps in three \textsc{gasoline2} simulations compared to data from \citet{Corbelli2014}.}
\label{fig:gasolinestars}
\end{figure}

The density of all stars within the 1\%, 10\% and 20\% feedback models is shown in Fig\;\ref{fig:gasolinestars}. This is compared to the stellar map of \citet{Corbelli2014}, see also the M33 image in the 2MASS Large Galaxy Atlas. As already noted, there is very little structure seen in the observed data, save for a few hints of spiral arms (most noticeably just to the south-south-east of galactic centre in the mid disc). All stars, both those formed in the simulation and those present in the initial conditions, are used for this figure. Generally the stellar maps show only small differences in structure, and thus the stellar map is not so useful to distinguish between the different models. The lowest feedback model has the greatest degree of structure, with clear spiral arms both in the inner and outer disc. The highest feedback model shows the best agreement with the M33 data, with very little consistent spiral arm features in the stars. The feedback strength keeps the gas disc dynamically hot enough that it prevents the growth of stellar spiral features. In this case the $Q$ or $m$ swing parameters are poor indicators of the stability of the stellar disc, as both should effectively be the same for all of the simulations shown, though neglect to take into account any gas physics.
\begin{figure}
\centerline{\includegraphics[scale=0.15]{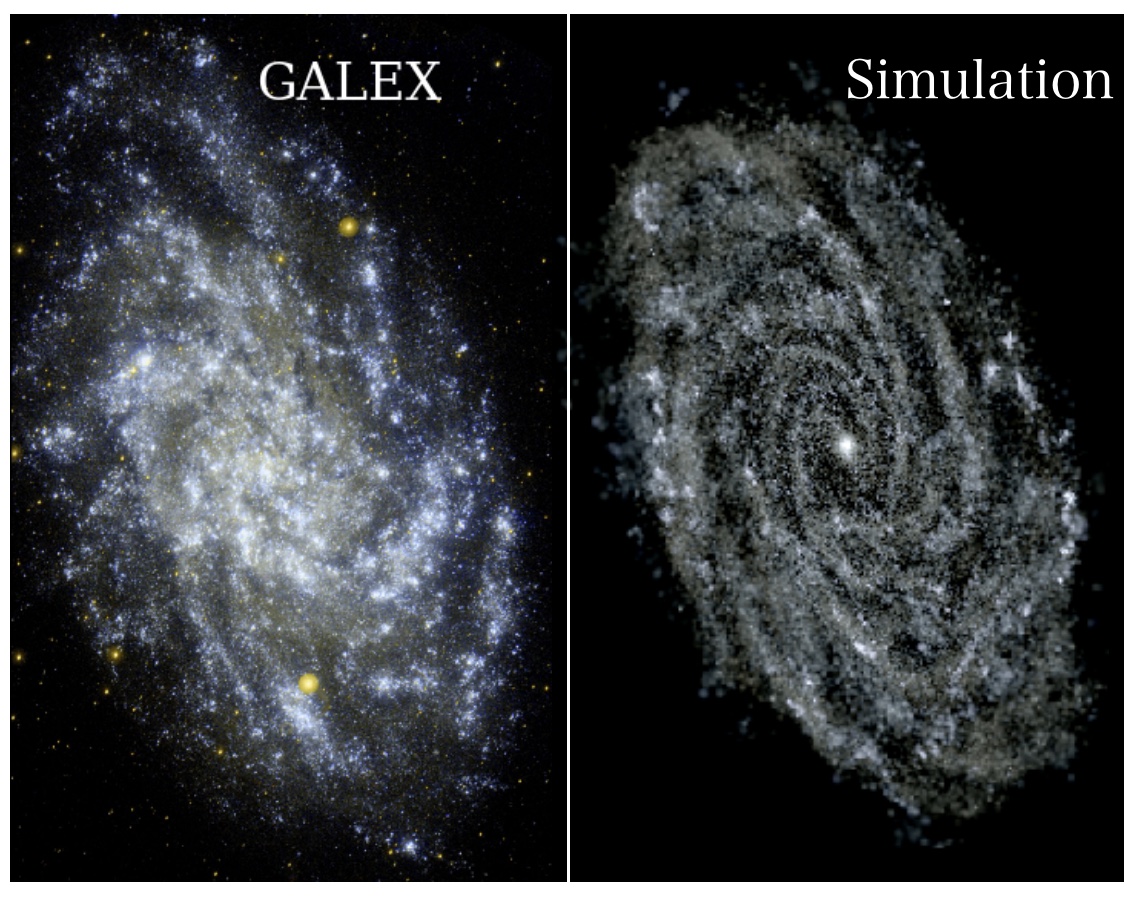}}
\caption{Left: GALEX UV map of M33, right: mock stellar image of the stars formed in the 10\% feedback \textsc{gasoline2} simulation.}
\label{fig:youngstars}
\end{figure}

We show the a mock stellar map of the young stellar population in the 10\% calculation (at the same time and orientation as used in the previous plots) in Figure\;\ref{fig:youngstars}. The image was created using the ages and masses of the young star particles and the \textsc{pynbody} software package (no dust attenuation is included). Young stars are simply defined as any stars not present in the initial conditions, i.e. those that have formed from gas particles throughout the simulation. We use the GALEX (credit: NASA/JPL-Caltech) UV map as a rough proxy for the young stellar structures in our simulated map. We only show a single map, as the 20\% and 10\% show little difference, whilst the 1\% and 5\% models are already deemed poor matches from the gas analysis.
The young star formation regions in the simulation show a general similarity to the UV map, with filamentary and patchy pockets of young stars. The main inconsistency is the inner disc, where there is a dearth of young stars at this point in the simulation. Although earlier frames are in a little better agreement, both the young stars and $\Sigma_{\rm SFR}$ do not reach the same high level seen in observations probably because we do not include molecular gas production, and as such SFR is a product of dense atomic rather than molecular gas content.

\section{Comparison of \textsc{sphNG} and \textsc{gasoline2} models}
\subsection{Star formation rates}
\begin{figure}
\centerline{\includegraphics[scale=0.5]{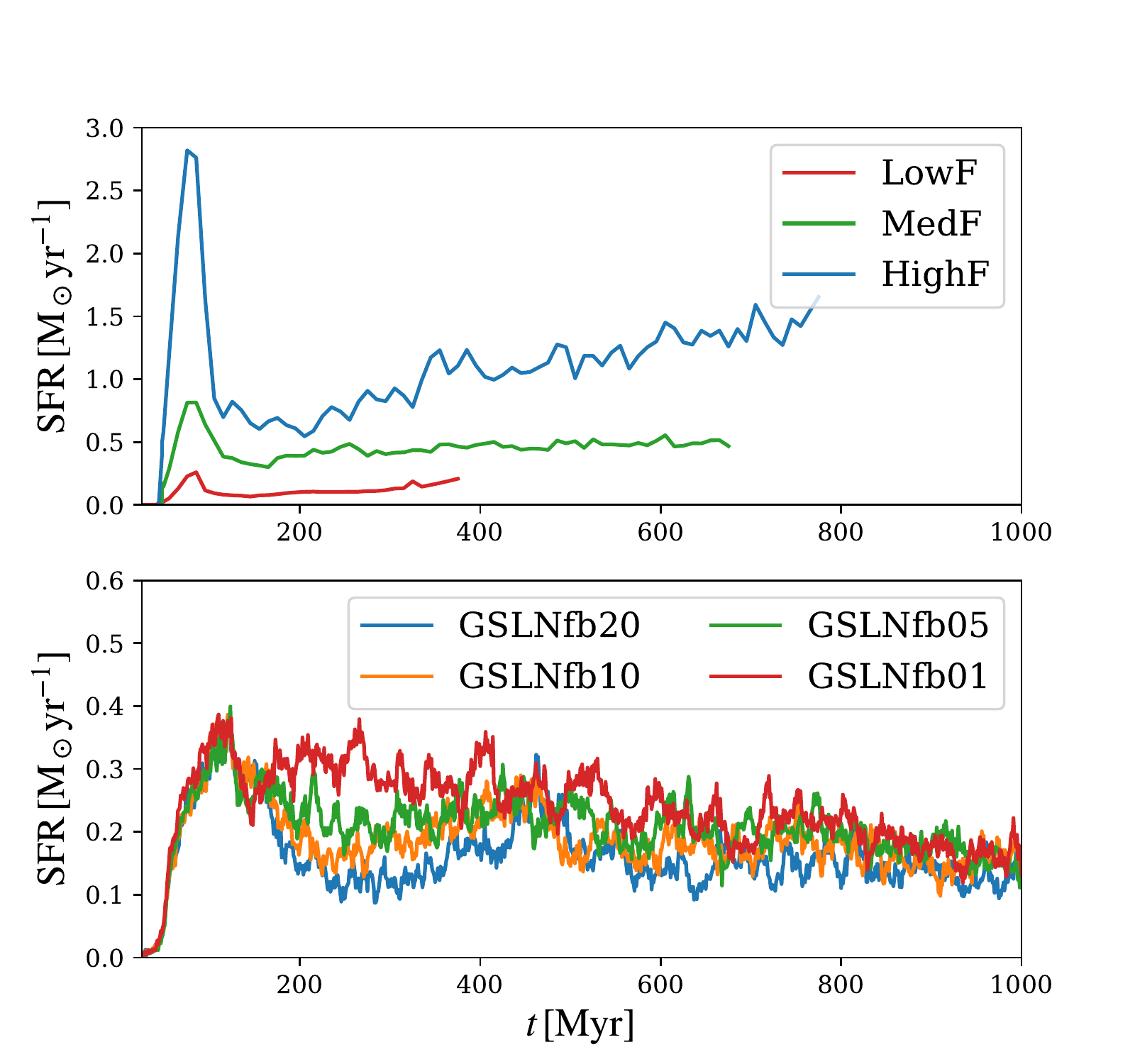}}
\caption{Star formation rate as a function of time in all \textsc{sphNG} calculations (top) and for the \textsc{gasoline2} calculations with different levels of feedback (lower).}
\label{fig:SFgasoline}
\end{figure}

In this Section we  compare the star formation rates from the \textsc{sphNG} and \textsc{gasoline2} models with different feedback. Fig\;\ref{fig:SFgasoline} shows the star formation rate versus time for the different models. The star formation rates clearly tend to be higher and more varied in the \textsc{sphNG} models. This is mainly because in these models, the star formation rate and feedback are controlled by a single parameter, whereas in the \textsc{gasoline2} models, there is both a star formation efficiency (constant) and a feedback efficiency (which varies) hence the star formation rates are more similar. Both sets of models show initial bursts of star formation, particularly in the higher feedback models, after which the star formation rate is comparatively more steady. With low feedback, there is less decrease (in the \textsc{gasoline2} model, no decrease) in the star formation rate after the initiation of star formation. Differences in the star formation rate, and the amount of energy added to the interstellar medium, likely explain the difference between the highest feedback models with the different codes. The MedF \text{sphNG} model appears similar to the \textsc{gasoline2} models and exhibits a similar star formation rate. The observed star formation rate in M33 is around 0.5 M$_{\odot}$ yr$^{-1}$ \citep{Verley2009}, which is  slightly higher than the \textsc{gasoline2} models, and comparable to the MedF \textsc{sphNG} model. Our best fit \textsc{sphNG} model, HighF, produces a higher star formation rate than is currently observed for M33.

\begin{figure}
\centerline{\includegraphics[scale=0.5]{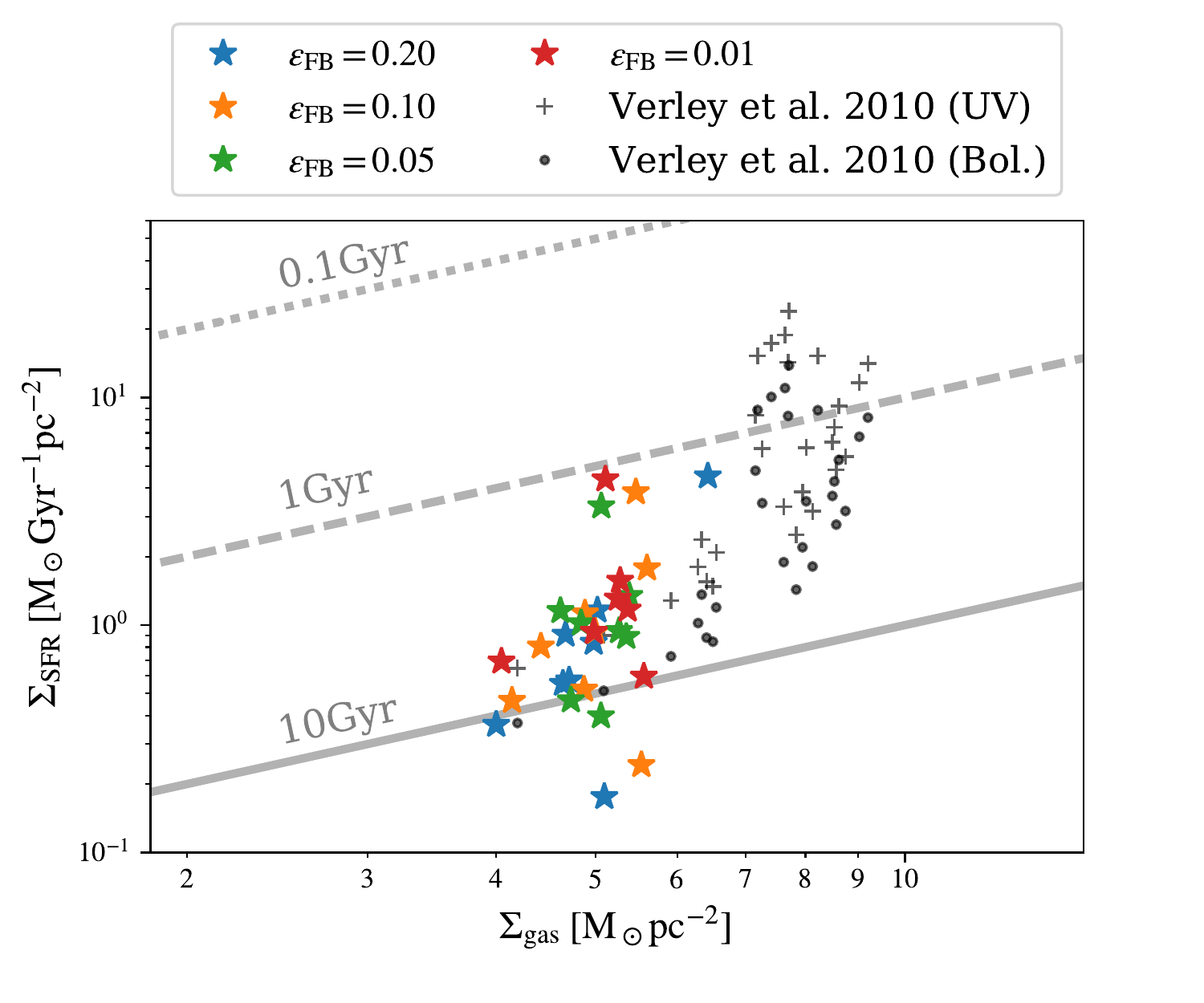}}
\caption{Surface star formation plotted against gas surface density in the \textsc{gasoline2} simulations after 1Gyr of evolution. FUV and bolometrically derived surface star formation rates are plotted versus atomic gas data for M33 from \citet{Verley2010}, shown by black dotted and plus points. Grey lines indicate constant depletion timescales.}\label{fig:KSgasoline}
\end{figure}

The surface star formation rate, $\Sigma_{\rm SFR}$, is shown in Figure\;\ref{fig:KSgasoline} for the \textsc{gasoline2} runs as a function of gas surface density, $\Sigma_{\rm gas}$. The data is binned up into annuli of width 1kpc moving from the centre to 7 kpc beyond which the surface density drops rapidly. Data is shown after 1Gyr of evolution, though different time-stamps showed no discernible difference. Also shown is the observational data from \citet{Verley2010}, where we have selected their atomic gas data plotted against bolometric and FUV surface star formation rates. All models trace effectively the same region of parameter space, and produce a good match to the observational data terms of dispersion. The gas surface densities are somewhat lower though compared to the data, with the \citet{Verley2010} data tracing surface densities higher on average than used to initialize our simulations which were constrained to the VLT+GBT data of \citet{Corbelli2014}. Also note that the gas budget of the \textsc{gasoline2} simulations is continuously being depleted over time by star formation, as well as ejected out of plane compared to the idealised initial conditions via feedback. As such, after 1Gyr the radially averaged surface density will be to some extent lower than the initial values, despite feedback continually delivering mass back from stars into the local ISM. The star formation rate is also slightly lower (as seen in Fig.\;\ref{fig:SFgasoline}), though some points to reach up to the 1Gyr depletion timescales as seen in the \citet{Verley2010} data. This deficit in star formation be an effect of resolution or simply insufficient star formation efficiency; a parameter that was fixed to a fiducial 10\% value for all \textsc{gasoline2} calculations. 

\subsection{The velocity field}
\begin{figure}
\centerline{\includegraphics[bb=150 300 1000 1500, scale=0.24]{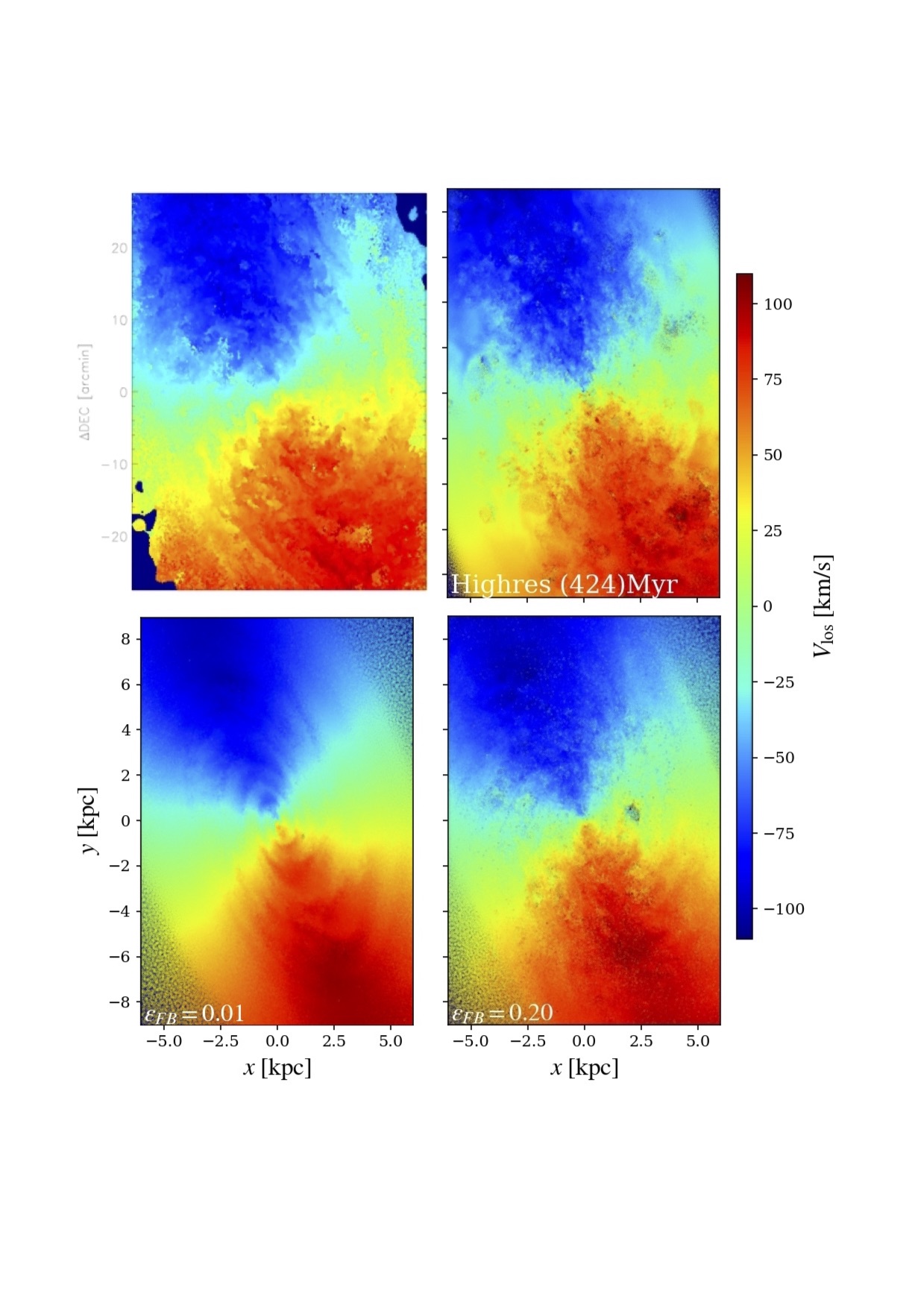}}
\caption{Velocity map from \citet{Corbelli2014} (top left) compared to the Highres \textsc{sphNG} simulation and \textsc{gasoline2} simulations with 1\% and 20\% feedback. The velocity-scale is the same for observations and simulation data (ignoring the 180km/s recession velocity for M33). Maps for the \textsc{gasoline2} simulations with 1\% and 20\% feedback are shown at the best-fitting time-frames of Fig.\;\ref{fig:gasolineside}.}\label{fig:vmapgasoline}
\end{figure}
We now compare our simulation data to the gas velocity field of M33, to further identify whether a high or low feedback recipe is the better match. Figure\;\ref{fig:vmapgasoline} shows the velocity field map of \citet{Corbelli2014} in the upper left compared to the gas velocity field of the 1\% \textsc{gasoline2} model and the 20\% feedback models from both codes. The \textsc{gasoline2} plots are at the same time and orientation shown in Figure\;\ref{fig:gasolineside}. The line-of-sight velocity is shown from the reference of the recession velocity of M33. 

The observational velocity field data shows many small scale features, with clear departures from pure circular rotation throughout. The lowest feedback model has clear shortcomings here, showing a very smooth velocity field structure. The few non-circular motions are seen around the spiral arms, and appear almost as concentric rings rather than arms. The higher feedback models show an excellent reproduction of the features seen in M33, with departures from circular rotation appearing in a very similar manner.

\section{Conclusions}
We have performed simulations of an isolated galaxy set up with similar gas, stellar and dark matter profiles to M33. Calculations were performed with two different SPH codes, sphNG and Gasoline2, with broadly consistent results between the two. We find that our models can reasonably reproduce the gaseous spiral structure of M33. The spiral features form as a result of gravitational instabilities in the stars and gas. Most previous work which has tried to explain spiral patterns in specific galaxies has concentrated on the generation of two armed spiral patterns by interactions with companions. Previous works however have not tried to reproduce a specific observed non-grand design galaxy. Our results thus indicate that a perturbing galaxy is not needed to produce spiral arms in M33, and that gravitational instabilities alone reproduce the observed spiral structure. 

Our results suggest that the gas is very relevant to the formation of the spiral arms, as well as the stars. We see very different morphologies in the gas and stars depending on the gas physics. In particular with a low feedback efficiency, strong spiral arms form in both the stars and gas, whereas weaker stellar arms, and more fragmented gaseous arms occur with a higher level of feedback. We suggest that the presence of gas allows the dissipation of energy in the stellar spirals, lowering $Q$ in the stars and maintaining the spiral arms for longer, although most of our simulations start with $Q\gtrsim1$ and do not show a large increase in $Q$ likely because the disc isn't so unstable and therefore the increase in $Q$ is only very moderate. We only see a difference in $Q$ when gas is included, compared to modelling only the stars, when $Q<1$ and the disc is more strongly unstable. Although we did not see a difference in $Q$ for models with different feedback efficiencies, we noted that there is more cold gas at lower efficiencies, which means the combined $Q$ for the gas and stars is smaller compared to the other simulations. This likely leads to the production of very strong spiral features in the model with the lowest levels of feedback (LowF). 

We find that best agreement with M33 occurs with a high level of stellar feedback, in agreement with the findings of \citet{Rahimi2012}. This is because the higher feedback blows out large holes in the gas creating clear voids which resemble the holes in M33, and because the gas is pushed into a smaller number of more prominent spiral arms. With a lower level of feedback, the gas is disrupted everywhere on small scales, producing many small fragments. With a very low feedback level, we get clear spiral arms, but they are much brighter than the actual M33, and appear continuous whereas the actual M33 does not contain long continuous spiral arms. For the feedback schemes in the codes presented here, we find good agreement with a feedback efficiency of 10--20 \%, though more generally this value may vary according to the precise details of the feedback implemented. The stellar discs maintain a relatively feature-less morphology in the higher feedback calculations, which is in agreement with observed stellar mass maps of M33, displaying only faint spiral arm features in the mid/outer disc.

We checked the validity of our results by performing simulations with different resolutions, and with the two different codes. The agreement with different resolution, and the dependence on the structure with feedback, is encouraging. Both show that intriguingly the worst agreement with M33 occurs with moderate levels of feedback (around 5\%), since the large voids between the spiral arms seen in M33 are only reproduced by large amounts of feedback (10\%--20\%), or in the case of low feedback stronger amplitude spiral arms occur in the stellar disk which gather gas into many elongated filaments or arms. We do find some difference in the dependence of feedback between the codes. In the gasoline models the size of the voids and arm segments are less sensitive to the feedback efficiency compared to the models using \textsc{sphNG}. This is not unreasonable given they are two different codes with different feedback schemes, gravity pipelines, cooling functions, star formation recipes, etc. 

Finally we show a number of observational comparisons to M33. Maps of the line-of-sight velocity in the gas shows a good agreement with M33 only in the case of high levels of feedback. Such feedback creates departures from circular rotation as seen in M33, whereas low and medium levels tend to show much smoother features. The star formation rates and efficiencies are in reasonable agreement with observed data, though neither is a perfect match, with \textsc{gasoline2} models being systematically slightly lower than observed values, and \textsc{sphNG} values varying with different levels of feedback, though the medium level of feedback produces a remarkable agreement with the observed value.

In this paper we have focused on the large scale structure of M33, and shown very good agreement with the observed spiral structure of M33. Our models also provide the opportunity to investigate the giant molecular cloud population of M33, and the links between molecular clouds and star formation in one of nearest neighbour galaxies, a topic we will consider in future work.

\section*{Acknowledgments}
We would like to thank the referee for a useful report which improved the presentation of this paper.
Calculations for this paper were performed on the DiRAC machine `Complexity', and the ISCA High Performance Computing Service at the University of Exeter. CLD acknowledges funding from the European Research Council for the FP7 ERC starting grant project LOCALSTAR. Numerical computations were also carried out on Cray XC30 at Center for Computational Astrophysics, National Astronomical Observatory of Japan.
We utilised the \textsc{pynbody} \citep{2013ascl.soft05002P} for post-processing and analysing of the \textsc{tipsy} files created by \textsc{gasoline2}. 
Several figures in this paper were produced using \textsc{splash} \citep{splash2007}.

\bibliographystyle{mn2e}
\bibliography{Dobbs}

\bsp
\label{lastpage}

\end{document}